\documentclass[prd,aps,preprint,tightenlines,superscriptaddress,nofootinbib,showpacs,showkeys,floatfix]{revtex4}
\usepackage{amsfonts}
\usepackage{array}
\usepackage{epsfig}
\usepackage{rotating}
\usepackage{multirow}

\usepackage{amsmath}    
\usepackage{verbatim}   
\usepackage{color}      
\usepackage{colordvi}
\usepackage{hyperref}   

\usepackage{graphicx}

\def\permille{\ensuremath{{}^\text{{\it o}}\mkern-5mu/\mkern-3mu_\text{{\it oo}}}}

\graphicspath{{../pics/}}
\DeclareGraphicsExtensions{.eps,.ps}

\usepackage{pslatex}
\usepackage{txfonts}
\usepackage[latin1]{inputenc}
\usepackage[T1]{fontenc}

\begin{document}
\begin{titlepage}
\thispagestyle{empty}
\noindent
DESY 21-002\\
\hfill
March 2021 \\
\vspace{1.0cm}

\begin{center}
  {\bf \Large
    Precision studies for Drell-Yan processes at NNLO
  }
  \vspace{1.25cm}

 {\large
   S.~Alekhin$^{\, a,b}$, 
   A.~Kardos$^{\, c,d}$, 
   S.~Moch$^{\, a,d}$ 
   and
   Z.~Tr\'ocs\'anyi$^{\, e,d}$
   \\
 }
 \vspace{1.25cm}
 {\it
   $^a$ II. Institut f\"ur Theoretische Physik, Universit\"at Hamburg \\
   Luruper Chaussee 149, D--22761 Hamburg, Germany \\
   \vspace{0.2cm}
   $^b$Institute for High Energy Physics \\
   142281 Protvino, Moscow region, Russia\\
   \vspace{0.2cm}
$^c$Department of Experimental Physics, University of Debrecen \\
   4010 Debrecen, PO Box 105, Hungary\\
   \vspace{0.2cm}
   $^d$ELKH-DE Particle Physics Research Group, University of Debrecen \\ 
   4010 Debrecen, PO Box 105, Hungary\\
   \vspace{0.2cm}
   $^e$
   Institute for Theoretical Physics, ELTE E\"otv\"os Lor\'and University\\
   P\'azm\'any P\'eter 1/A, H--1117 Budapest, Hungary\\
 }
  \vspace{1.4cm}
  \large {\bf Abstract}
  \vspace{-0.2cm}
\end{center}
We present a detailed comparison of the fixed-order predictions computed by four publicly available computer codes for Drell-Yan processes at the LHC and Tevatron colliders. We point out that while there is agreement among the predictions at the next-to-leading order accuracy, the predictions at the next-to-next-to-leading order (NNLO) differ, whose extent depends on the observable. The sizes of the differences in general are at least similar, sometimes larger than the sizes of the NNLO corrections themselves. We demonstrate that the neglected power corrections by the codes that use global slicing methods for the regularization of double real emissions can be the source of the differences. Depending on the fiducial cuts, those power corrections become linear, hence enhanced as compared to quadratic ones that are considered standard.
\end{titlepage}

\newpage
\setcounter{footnote}{0}
\setcounter{page}{1}

\section{Introduction}
\label{sec:intro}

The very high quality of data for many Standard Model (SM) scattering
processes collected at the Large Hadron Collider in recent years makes it
mandatory to use high precision theoretical predictions for physics analyses of these data.
This is true especially for the vector-boson hadroproduction that is the prime benchmark process at hadron colliders.
The theory computations need to account for radiative corrections,
the dominating ones being due to quantum chromodynamics (QCD), 
where presently the next-to-next-to-leading order (NNLO) is the state-of-the-art~\cite{Heinrich:2020ybq}.
In addition, the experimental cuts such as limits on the transverse
momenta or the rapidities of the observed final state particles 
applied during the data selection have to be included in the theory calculations. 
Cross section predictions have to be provided for the respective fiducial regions.

For colorless final states such as the Drell-Yan process or more specifically, 
for $W^{\pm}$- and $Z$-boson production including their decay, the QCD predictions 
are known to NNLO for fully exclusive kinematics~\cite{Catani:2007vq,Catani:2009sm,Gavin:2010az,Boughezal:2016wmq}.
These computations require the combination of squared matrix elements with three
different multiplicities of partons in the final state and the subsequent 
cancellation of the soft and collinear singularities upon integration over
their phase space to arrive at infrared finite results, 
a step which is performed with the help of dedicated subtraction schemes.
In summary, the problem is considered being solved and the NNLO QCD predictions 
for $W^{\pm}$- and $Z$-boson production including the leptonic decay 
have been made available in several computer programs 
which perform the integration of the parton level predictions numerically by Monte Carlo methods.
It has become apparent, however, that the level of accuracy of the published codes 
is not sufficient for the needs in analyses of experimental data. 
Significant deviations between the predictions of the different codes have
been documented, for example, in an ATLAS analysis~\cite{Aaboud:2016btc}.

The motivation for the present study comes from the use of data on
$W^{\pm}$- and $Z$-boson hadro-production collected at the LHC and the Tevatron 
in the determination of parton distribution functions (PDFs) at NNLO accuracy. 
Those data consist of differential distributions in the pseudo-rapidity of the
decay leptons and typically have a precision of ${\cal O}(1-2\%)$, which is mainly
dominated by the experimental systematics. 
This fact and the generally rather small size of the pure NNLO QCD corrections 
in the range of ${\cal O}(1\%)$ or even less relative to the fiducial cross
sections at next-to-leading order (NLO) lead us to investigate 
the precision of available QCD predictions. 
To that end, we focus on two aspects in this work.
For one, we provide benchmark numbers for NNLO QCD predictions in
kinematics which are representative for the bulk of the available experimental data. 
As a target we aim at predictions with a residual uncertainty 
of less than ${\cal O}(0.1\permille)$ in each bin from the Monte Carlo integration for the cross section integrated over the fiducial region,
whenever possible in order to have accurate results when comparing the different codes.
Previous comparisons~\cite{Alioli:2016fum} of some of the published codes 
have limitations in the precision of the predictions.
A second point concerns the investigation of the observed deviations 
among the codes in the light of the subtraction schemes used, 
which are either local or global, depending on whether the cancellations of the 
infrared singularities are performed locally in the integrand 
at each point in phase space or whether they are accomplished globally after 
integrating over a slice of the phase space.
In particular, we illustrate the impact of fiducial cuts on the
decay leptons for subtraction schemes with slicing.

The paper is organized as follows. 
In Sec.~\ref{sec:benchmark} we present the benchmark numbers for two 
representative sets of $W^{\pm}$- and $Z$-boson data from the LHC and the
Tevatron. We first validate the different codes at NLO in QCD and then quantify the
deviations at NNLO.
In Sec.~\ref{sec:powercorrections} we provide a brief review of global slicing
methods and the power counting in the slicing parameter. 
Then we compute the effect and the size of power corrections on the example of the lepton
decay phase space for the fiducial cuts of the LHC and Tevatron data considered
in the previous section. 
We conclude in Sec.~\ref{sec:conclusions}.

\section{Benchmark computations}
\label{sec:benchmark}

\subsection{Set-up and validation}

The set-up for benchmarking available QCD predictions for $W^{\pm}$- and
$Z$-boson hadro-production cross sections up to NNLO in QCD in the fiducial
phase space of the experimental measurement contains three aspects: 
the choice of the data sets, the list of input parameters and 
the selection of the NNLO QCD codes for the comparison of the theoretical
predictions. 

We choose two sets of data on $W^{\pm}$- and $Z$-boson production collected by
the ATLAS experiment at the LHC and the D{\O} experiment at the Tevatron, respectively, 
which are statistically significant in current fits of PDFs.
\begin{itemize}
\item 
  The ATLAS data set for the $W^\pm$- and $Z/\gamma^*$-production cross
  sections \cite{Aaboud:2016btc} measured at a center-of-mass energy of
  $\sqrt{s}=7$~TeV at the LHC.
  These data are given in form of pseudo-rapidity distributions for the decay electron or muon ($W^{\pm}$-production)
  and the decay lepton-pair ($Z/\gamma^*$-production), respectively.
  The transverse momenta $p_T^l$ and the pseudo-rapidities $\eta_l$ of the decay leptons are subject to fiducial cuts.
  The cross sections for $Z/\gamma^*$-production are measured at central as
  well as at forward pseudo-rapidities.

\item The data obtained by D{\O} on $W^\pm$-boson production at $\sqrt{s}=1.96$~TeV at the 
  Tevatron~\cite{D0:2014kma} measures the electron charge 
  asymmetry distributions and their dependence on the electron pseudo-rapidity. 
  These data also probe forward kinematics. 
  Also, the D{\O} data apply fiducial cuts, both symmetric as well as staggered, on the transverse momenta
  $p_T^{l,\nu}$ of the electron and the neutrino and on their pseudo-rapidities.
\end{itemize}

Another data set by ATLAS, the measurement of the muon charge asymmetry in
$W^{\pm}$-boson production at $\sqrt{s}=8$~TeV at the LHC~\cite{Aad:2019rou}
has similar experimental precision as the chosen data set~\cite{Aaboud:2016btc} collected
at $\sqrt{s}=7$~TeV and also largely overlaps in kinematics.
Similar considerations apply to data from CMS and LHCb, e.g.,~\cite{Khachatryan:2016pev,Aaij:2015zlq}
Hence, we do not include these data in the benchmark comparison.

We use the $G_\mu$ scheme with input values $G_F$, $M_Z$, $M_W$ and with 
$\sin^2(\theta_W)$ and the QED coupling $\alpha(M_Z)$ as output values. 
This scheme minimizes the impact of NLO electroweak corrections, see e.g. \cite{Dittmaier:2009cr}.
In detail, our SM input parameters are~\cite{pdg2020}
\begin{equation}
    \begin{array}{ll}
\label{eq:input-masses}
~~G_{\mu} = 1.16637\times 10^{-5} \; {\rm GeV}^{-2} \, , 
\\ 
~M_Z = 91.1876 \; {\rm GeV}\, , \quad\quad   &~\Gamma_{Z} =  2.4952  \; {\rm GeV} \, , 
\\ 
M_W = 80.379\; {\rm GeV}\, , \quad\quad  
&\Gamma_W = 2.085\; {\rm GeV} \, ,
\end{array}
\end{equation}
and for the relevant CKM parameters
\begin{eqnarray}
\label{eq:input-ckm}
|V_{ud}| \,=\, 0.97401\, , & \quad & |V_{us}| \,=\, 0.2265\phantom{0}  \, ,
\nonumber \\
|V_{cd}| \,=\, 0.2265\phantom{0} \, , & \quad & |V_{cs}| \,=\, 0.97320 \, ,
\nonumber \\
|V_{ub}| \,=\, 0.00361\, , & \quad & |V_{cb}| \,=\, 0.04053 \, .
\end{eqnarray}
The computations are performed in the $\overline{\rm MS}$ factorization scheme with $n_f=5$ light flavors. 
Therefore we take the $n_f=5$ flavor PDFs of ABMP16~\cite{Alekhin:2017kpj,Alekhin:2018pai} as an input 
together with the value of the strong coupling, $\alpha_s^{(n_f=5)}(M_Z) = 0.1147$. These choices do not bias any of the predictions we discuss below.
The renormalization and factorization scales $\mu_R$ and $\mu_F$ are set 
to $\mu_R = \mu_F = M_V$, where $M_V$ is the mass of the gauge boson $V$.

We run the following publicly available codes designed for 
the computation of the fully differential NNLO QCD predictions for the lepton
rapidity distributions.
\begin{itemize}
\item {\tt DYNNLO} (version 1.5)~\cite{Catani:2007vq,Catani:2009sm} 
  \footnote{Code available from \url{http://theory.fi.infn.it/grazzini/dy.html}.}
  (No built-in computation of PDF uncertainties available.)
  
\item {\tt FEWZ} (version 3.1)~\cite{Li:2012wna,Gavin:2012sy} 
  \footnote{Code available from \url{https://www.hep.anl.gov/fpetriello/FEWZ.html}.}
  
\item {\tt MATRIX} (version 1.0.4)~\cite{Grazzini:2017mhc}
  \footnote{Code available from \url{https://matrix.hepforge.org/}.}
  (No built-in computation of PDF uncertainties available.)\\
  {\tt MATRIX} uses the scattering amplitudes from {\tt OpenLoops} \cite{Cascioli:2011va}.
  
\item {\tt MCFM} (version 9.0)~\cite{Campbell:2019dru}
  \footnote{Code available from \url{https://mcfm.fnal.gov/}.}\\
  {\tt MCFM} uses the implementation of the NNLO computation of Ref.~\cite{Boughezal:2016wmq}.

\end{itemize}

The codes differ by the subtraction schemes used. {\tt FEWZ} uses sector
decomposition and employs a fully local subtraction scheme. 
{\tt DYNNLO} and {\tt MATRIX} both use $q_T$-subtraction~\cite{Catani:2007vq} 
at NNLO, which is a global phase space slicing method.
{\tt MCFM} uses $N$-jettiness subtraction~\cite{Boughezal:2015dva,Gaunt:2015pea}
at NNLO, which is also a global phase space slicing method, 
see \cite{TorresBobadilla:2020ekr} for a recent review.
Codes with global slicing do require a slicing parameter. 
These are $r_{\rm cut}$ for {\tt MATRIX} as a cut on $q_T$ and 
$\tau_{\rm cut}$ for {\tt MCFM} as the jettiness slicing parameter.

The {\tt DYNNLO} program is a legacy code, now superseded by {\tt MATRIX}. 
It has been the first publicly available program containing the NNLO QCD predictions 
for fully exclusive kinematics and it is included in this list because of 
the ATLAS study~\cite{Aaboud:2016btc} and its continued use in the analyses of experimental data. 
An improved reimplementation of the {\tt DYNNLO} code is also part of program
{\tt DYTurbo} for fast predictions for Drell-Yan
processes~\cite{Camarda:2019zyx} 
\footnote{Code available from \url{https://dyturbo.hepforge.org/}.}.
{\tt DYTurbo} includes the resummation of
large logarithmic corrections, too. In the fixed-order mode, software profiling 
was employed to achieve code optimization, but it reproduces exactly the
results of {\tt DYNNLO} within numerical uncertainties due to the different integration method, 
hence its predictions are not included in our benchmark comparisons.
Another code which we mention for reference is {\tt SHERPA-NNLO-FO}~\cite{Hoeche:2014aia} 
\footnote{Code available from \url{https://slac.stanford.edu/~shoeche/pub/nnlo/}.}.
We do not include this code in the benchmark exercise and instead refer to the
previous study~\cite{Alioli:2016fum}.

We start with the validation up to NLO of the codes selected for the comparison. 
This serves a two-fold purpose. 
First, it provides a check on the input settings for the codes and 
second, it demonstrates the level of agreement for the benchmark results. 
We show first the results for the predictions for the $W^\pm$- and 
$Z/\gamma^*$-production cross sections at $\sqrt{s}=7$~TeV corresponding to the kinematics of the 
ATLAS data set~\cite{Aaboud:2016btc}. 
In all cases, here and below we use the ABMP16 PDFs at NNLO and the
value $\alpha_s^{(n_f=5)}(M_Z) = 0.1147$, independent of the order of
perturbation theory for the cross sections $\sigma_{\rm LO}$, $\sigma_{\rm NLO}$ and $\sigma_{\rm NNLO}$.
We have found excellent agreement for all cross sections computed at the leading order (LO, not shown here) 
at ${\cal O}(10^{-5})$. 
At NLO we can directly compare and quantify the accuracy of the numerical integrations, 
as all codes employ fully local subtraction schemes except for {\tt DYNNLO},
which uses $q_T$-subtraction also at NLO~\footnote{
Dipole subtraction is used in {\tt DYNNLO} only in the computation of the vector-boson+jet contribution.}.
{\tt FEWZ} applies sector decomposition while {\tt MATRIX} and {\tt MCFM} 
all use by default the dipole subtraction~\cite{Catani:1996vz}.
\begin{figure}[t!]
\begin{center}
\includegraphics[width=8.1cm]{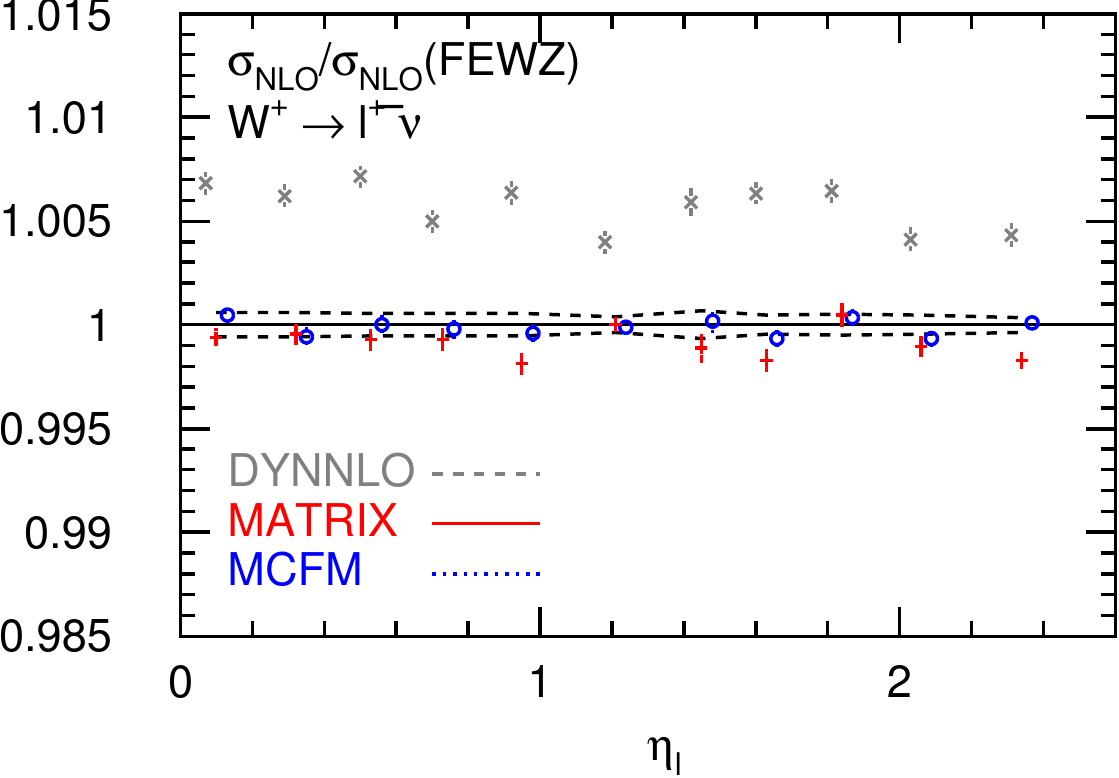}
\includegraphics[width=8.1cm]{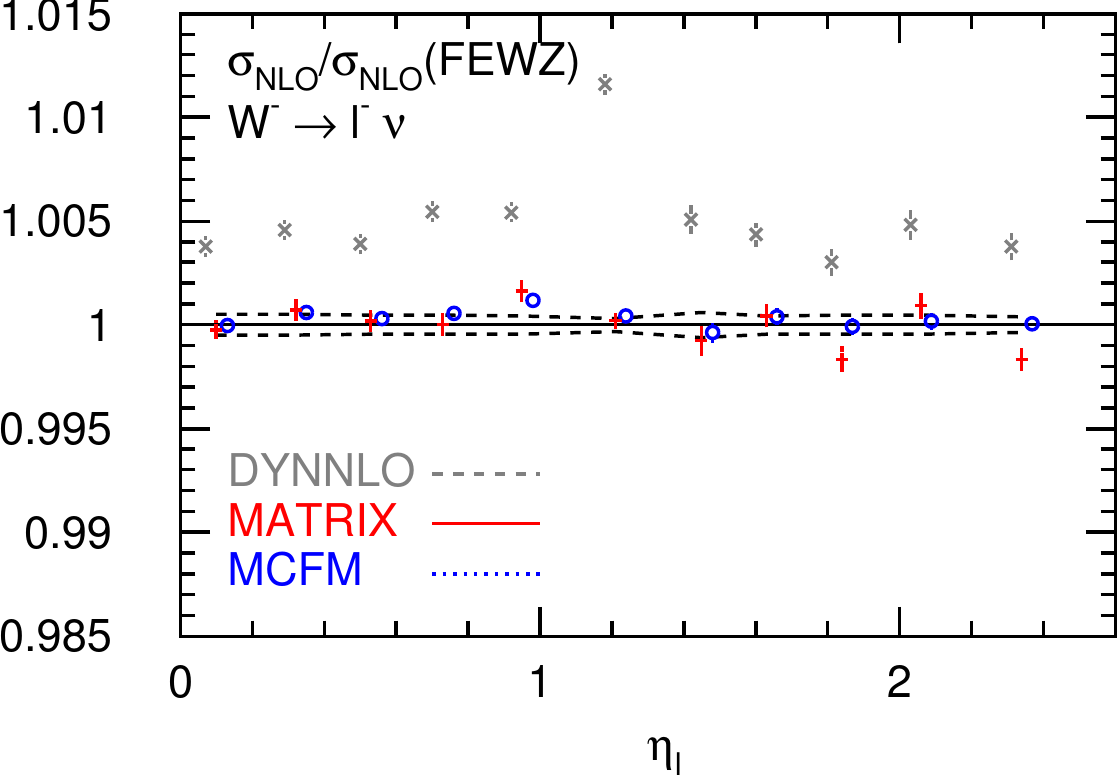}
\caption{\small
  \label{fig:atlas-Wpm-nlo}
  The NLO QCD cross sections for inclusive $pp \to W^\pm+X \to l^\pm \nu + X$ 
  as function of pseudo-rapidity $\eta_l$, 
  computed with {\tt DYNNLO}, {\tt MATRIX} and {\tt MCFM} relative to {\tt  FEWZ} 
  and using the ABMP16 PDFs.
  Cuts of $p_T^{l,\nu} \geq 25$\,GeV and $M_T \geq 40$\,GeV
  for the transverse momenta and mass are applied as in the  
  ATLAS data selection~\cite{Aaboud:2016btc}.
  The error bars indicate the accuracy of the numerical integration, 
  shown by the horizontal dashed lines for the {\tt FEWZ} result. 
  The {\tt MATRIX} result is plotted in the center of each $\eta_l$-bin, while 
  {\tt DYNNLO} and {\tt MCFM} results are shifted slightly to the left and the
  right.
}
\end{center}
\end{figure}
\begin{figure}[t!]
\begin{center}
\includegraphics[width=8.1cm]{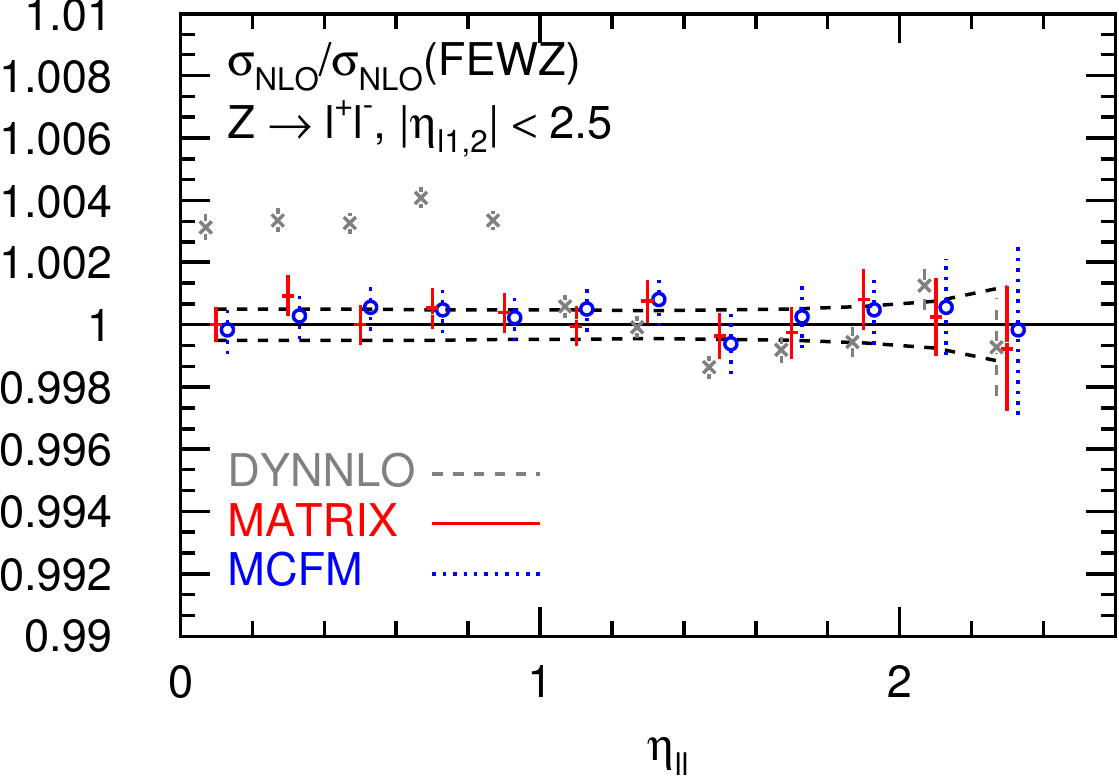}
\caption{\small
  \label{fig:atlas-Z-central-nlo}
  Same as Fig.~\ref{fig:atlas-Wpm-nlo} for the NLO QCD cross sections for 
  $pp \to Z/\gamma^*+X \to l^+l^- + X$ production at $\sqrt{s}=7$~TeV 
  as function of the pseudo-rapidity $\eta_{ll}$. 
  Cuts of $p_{T1,2} \geq 25$~GeV, $66 \geq M_{ll} \geq 116$~GeV and 
  $|\eta_{l_i}| \leq 2.5$, $i=1,2$ for the lepton pseudo-rapidities are applied.
}
\end{center}
\end{figure}
\begin{figure}[t!]
\begin{center}
\includegraphics[width=8.1cm]{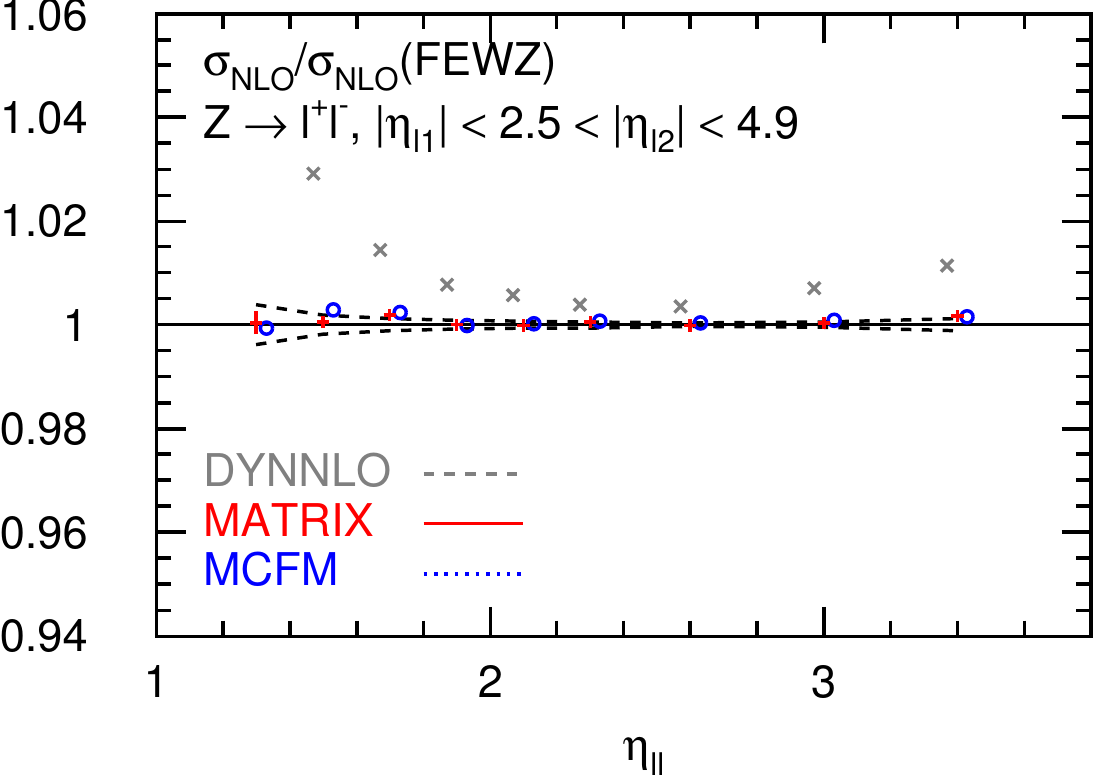}
\includegraphics[width=8.1cm]{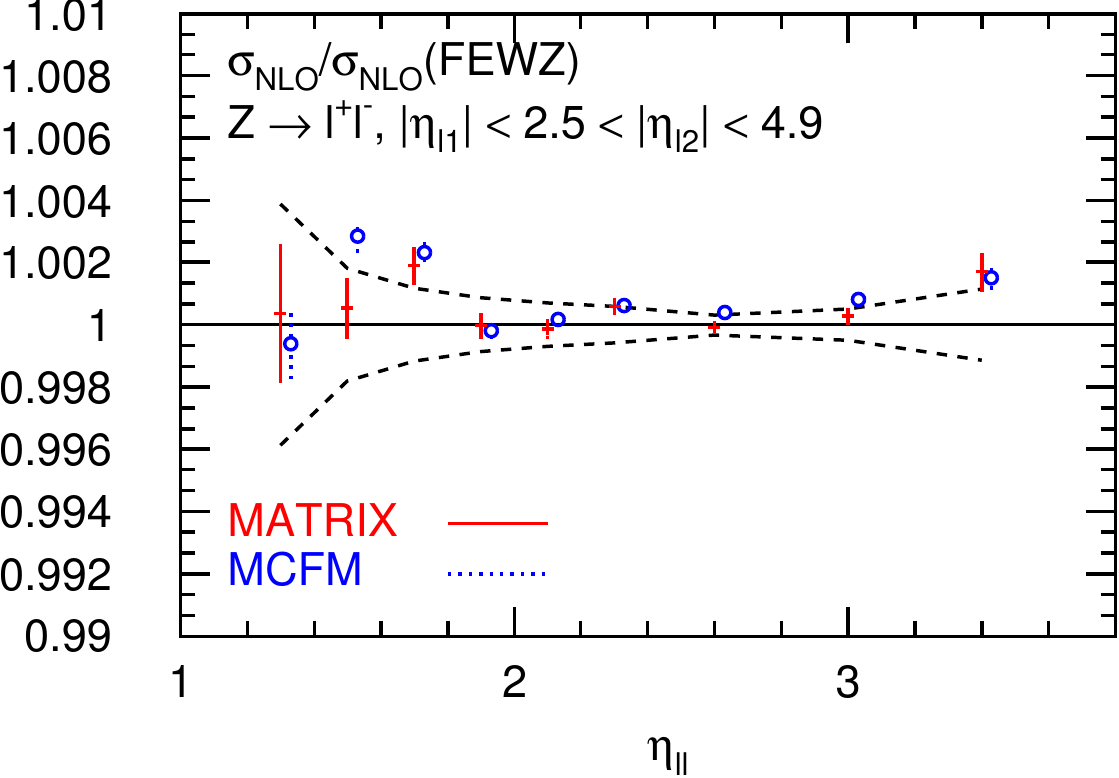}
\caption{\small
  \label{fig:atlas-Z-forward-nlo}
  Same as Fig.~\ref{fig:atlas-Z-central-nlo} with one lepton at forward 
  pseudo-rapidities and cuts of $|\eta_{l_1}| \leq 2.5$ and $2.5 \leq |\eta_{l_2}| \leq 4.9$.
  Predictions by {\tt DYNNLO}, {\tt MATRIX} and {\tt MCFM} relative to
  {\tt  FEWZ} (left) and zoom on {\tt MATRIX} and {\tt MCFM} results (right).
}
\end{center}
\end{figure}
\begin{figure}[t!]
\begin{center}
\includegraphics[width=8.1cm]{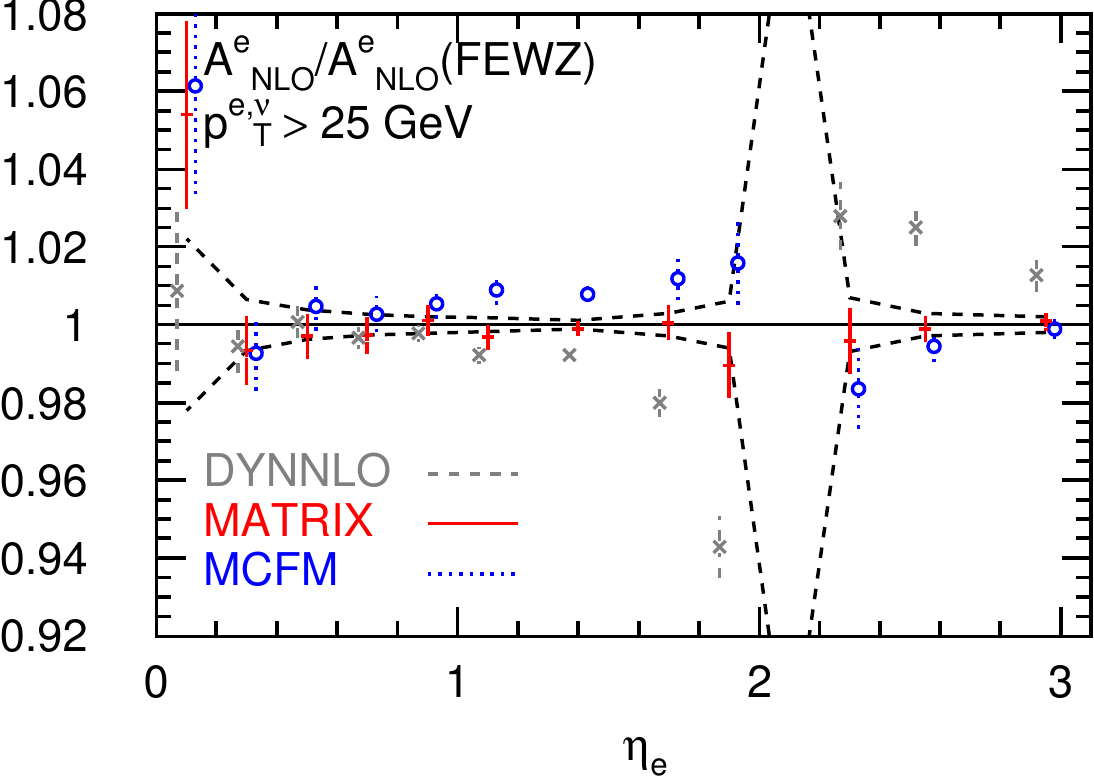}
\includegraphics[width=8.1cm]{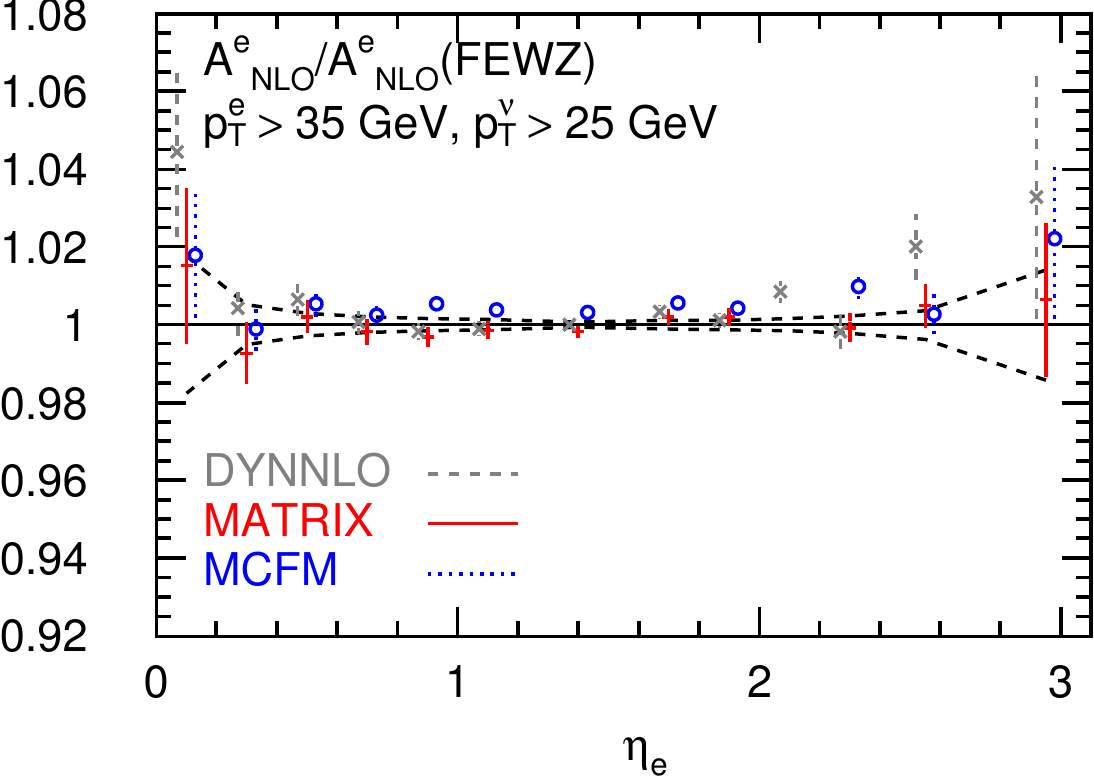}
\caption{\small
  \label{fig:d0-asy-nlo}
  The ratio of the NLO QCD corrections to the electron charge 
  asymmetry distribution $A_e$ in $W^{\pm}$-boson production at $\sqrt{s}=1.96$~TeV 
  computed with {\tt DYNNLO}, {\tt MATRIX} and {\tt MCFM} to ones by {\tt FEWZ}. 
  Cuts are applied as in the D{\O} data selection~\cite{D0:2014kma}: 
  $p_T^{e,\nu} \geq 25$~GeV symmetric (left) and 
  $p_T^{e} \geq 35$~GeV and $p_T^{\nu} \geq 25$~GeV staggered (right).
}
\end{center}
\end{figure}

In Fig.~\ref{fig:atlas-Wpm-nlo} we plot the results for the $W^\pm$-production 
cross sections at $\sqrt{s}=7$~TeV at NLO corresponding to the kinematics of the 
ATLAS data set~\cite{Aaboud:2016btc} and we find agreement at the level of ${\cal O}(1\permille)$ 
between {\tt FEWZ}, {\tt MATRIX} and {\tt MCFM} in the entire $\eta_l$ range,
while for {\tt DYNNLO} we find the values to be systematically enhanced by ${\cal O}(5\permille)$ 
for both, $W^+$- and $W^-$-production.
In Figs.~\ref{fig:atlas-Z-central-nlo} and \ref{fig:atlas-Z-forward-nlo} we 
show the NLO cross sections for $Z/\gamma^*$-production at $\sqrt{s}=7$~TeV in 
the ATLAS kinematics with the different selection cuts on the lepton pseudo-rapidities.
In the case of both leptons at central pseudo-rapidities $|\eta_{l_i}| \leq 2.5$ for $i=1,2$ 
in Fig.~\ref{fig:atlas-Z-central-nlo} we find agreement
among all codes, except for a slight systematic off-set of the {\tt DYNNLO} 
result by ${\cal O}(3\permille)$ for $\eta_{ll} \lesssim 1.0$.
In Fig.~\ref{fig:atlas-Z-forward-nlo} we display the case when one lepton is required at central
and the other one instead at forward pseudo-rapidity, $|\eta_{l_1}| \leq 2.5 \leq |\eta_{l_2}| \leq 4.9$.
We observe agreement at the level of ${\cal O}(1 - 2\permille)$ between {\tt FEWZ}, {\tt MATRIX} and {\tt MCFM}
as shown in Fig.~\ref{fig:atlas-Z-forward-nlo} on the right, 
while the {\tt DYNNLO} results turn out to be larger by up to a few per cent 
in the first $\eta_{ll}$ bins. 
Finally, in Fig.~\ref{fig:d0-asy-nlo} we plot the 
electron charge asymmetry distribution $A_e$ in $W^{\pm}$-boson production at $\sqrt{s}=1.96$~TeV
for two choices of cuts applied in the selection of the D{\O} data~\cite{D0:2014kma}: 
on the left we have symmetric cuts, $p_T^{e,\nu} \geq 25$~GeV, 
and on the right staggered cuts, $p_T^{e} \geq 35$~GeV and $p_T^{\nu} \geq 25$~GeV. 
We show the ratio of the {\tt DYNNLO}, {\tt MATRIX} and {\tt MCFM} results with {\tt FEWZ}. 
The level of agreement for $A_e$ is better than ${\cal O}(1\%)$, 
except in regions, where NLO predictions are very small. 
For symmetric cuts in Fig.~\ref{fig:d0-asy-nlo} on the left, this happens 
in the first bin, where the relative agreement deteriorates to a few per cent, 
and in the bin $\eta_{e} = 2.1$, where $A_e$ vanishes and the corresponding ratios have not been plotted.
For staggered cuts in Fig.~\ref{fig:d0-asy-nlo} on the right, this is observed
also in the first bin and in the region $\eta_{e} \gtrsim 2.5$.
The individual cross sections for $W^\pm$-production are computed at an accuracy of ${\cal O}(0.1\permille)$,
but in the asymmetry, one looses more than one order of magnitude in precision.

In summary, the validation shows that the predictions by {\tt MATRIX}, {\tt MCFM} and {\tt FEWZ} are all 
in very good agreement at NLO. 
{\tt DYNNLO} using $q_T$-subtraction also at NLO delivers results, which are 
typically accurate up to a few per mill and deviate in particular 
for distributions with challenging kinematics like in
Fig.~\ref{fig:atlas-Z-forward-nlo}, where agreement can only be reached at the level of a few per cent.
Moreover, the deviations of the {\tt DYNNLO} results from the ones by {\tt MATRIX}, {\tt MCFM} and {\tt FEWZ} display a particular pattern as a function of the (di-)lepton pseudo-rapidities $\eta_l (\eta_{ll})$, which will be addressed 
in Sec.~\ref{sec:powercorrections} below in the light of power corrections in the slicing parameter. 

\subsection{NNLO benchmark predictions}

We now proceed to the QCD predictions at NNLO accuracy, again starting with 
$W^\pm$- and $Z/\gamma^*$-production cross sections at $\sqrt{s}=7$~TeV 
as measured by ATLAS data set~\cite{Aaboud:2016btc}. 
As a baseline for the comparison, 
we have computed the NNLO QCD predictions with the ABMP16 PDFs~\cite{Alekhin:2017kpj} and {\tt FEWZ} 
as this is the only available code which implements a fully local subtraction scheme at NNLO. 
We emphasize that our conclusions do not depend on the choice for the default theoretical prediction.
Note, that the ATLAS data have been released after completion
of the ABMP16 PDFs and were not included in the fit.
However, these data are in a good agreement with the predictions.

\begin{figure}[t!]
\begin{center}
\includegraphics[width=16.5cm]{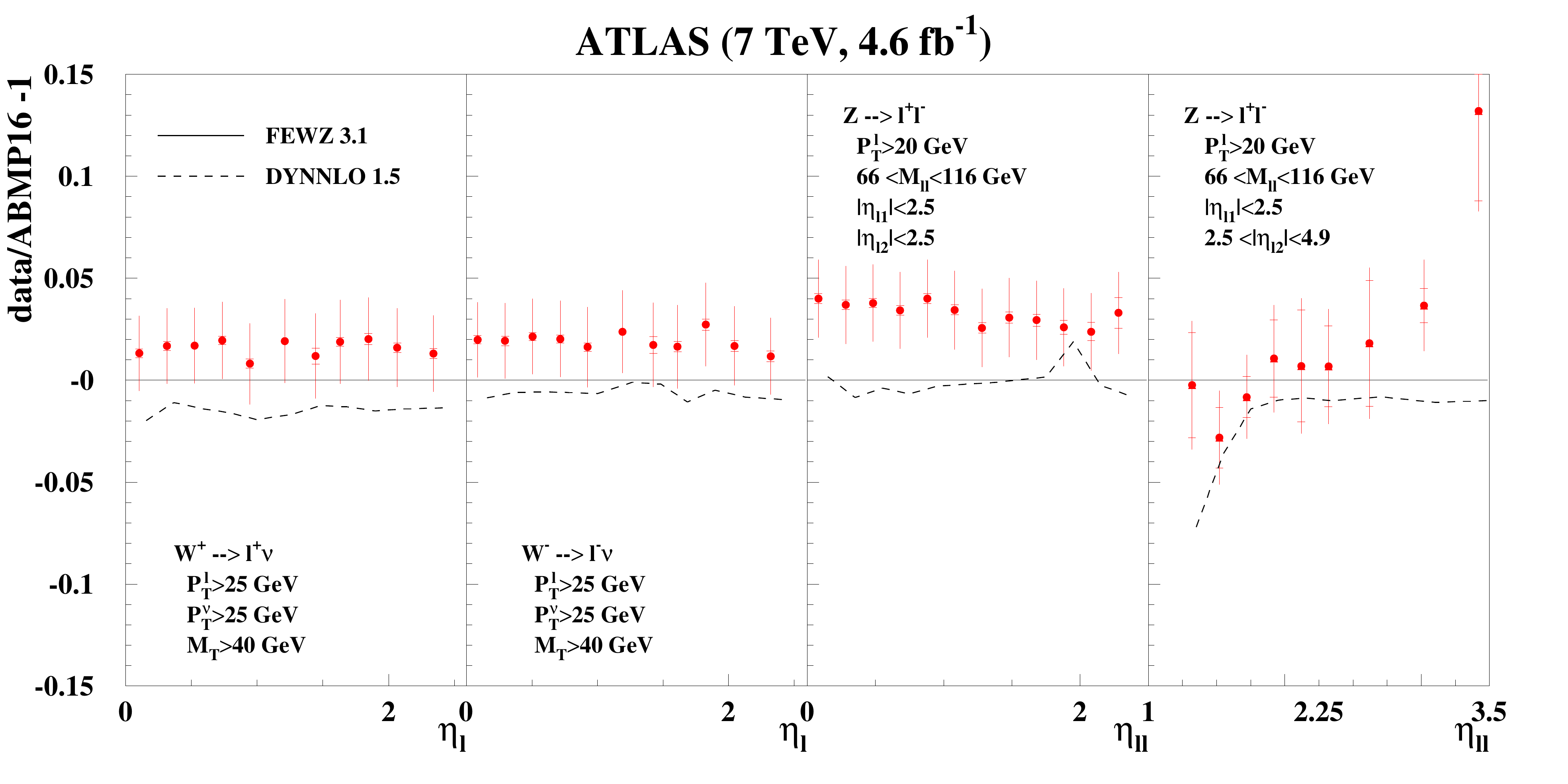}
\caption{\small
  \label{fig:dynnlo-atlas7data}
  The pulls for the ATLAS data measured 
  in inclusive $pp \to W^\pm+X \to l^\pm \nu + X$ 
  and $pp \to Z/\gamma^*+X \to l^+l^-  + X$ production 
  at $\sqrt{s}=7$~TeV~\cite{Aaboud:2016btc} with the 
  statistical (inner bar) and the total uncertainties, including the 
  systematic ones. The fiducial cuts on the decay leptons in the final state
  are indicated in the figure.
  The ABMP16 central predictions at NNLO are obtained
  with {\tt FEWZ} and the deviations of the predictions from {\tt DYNNLO} 
  are shown (dashed) for comparison. 
}
\end{center}
\end{figure}
\begin{figure}[t!]
\begin{center}
\includegraphics[width=16.5cm]{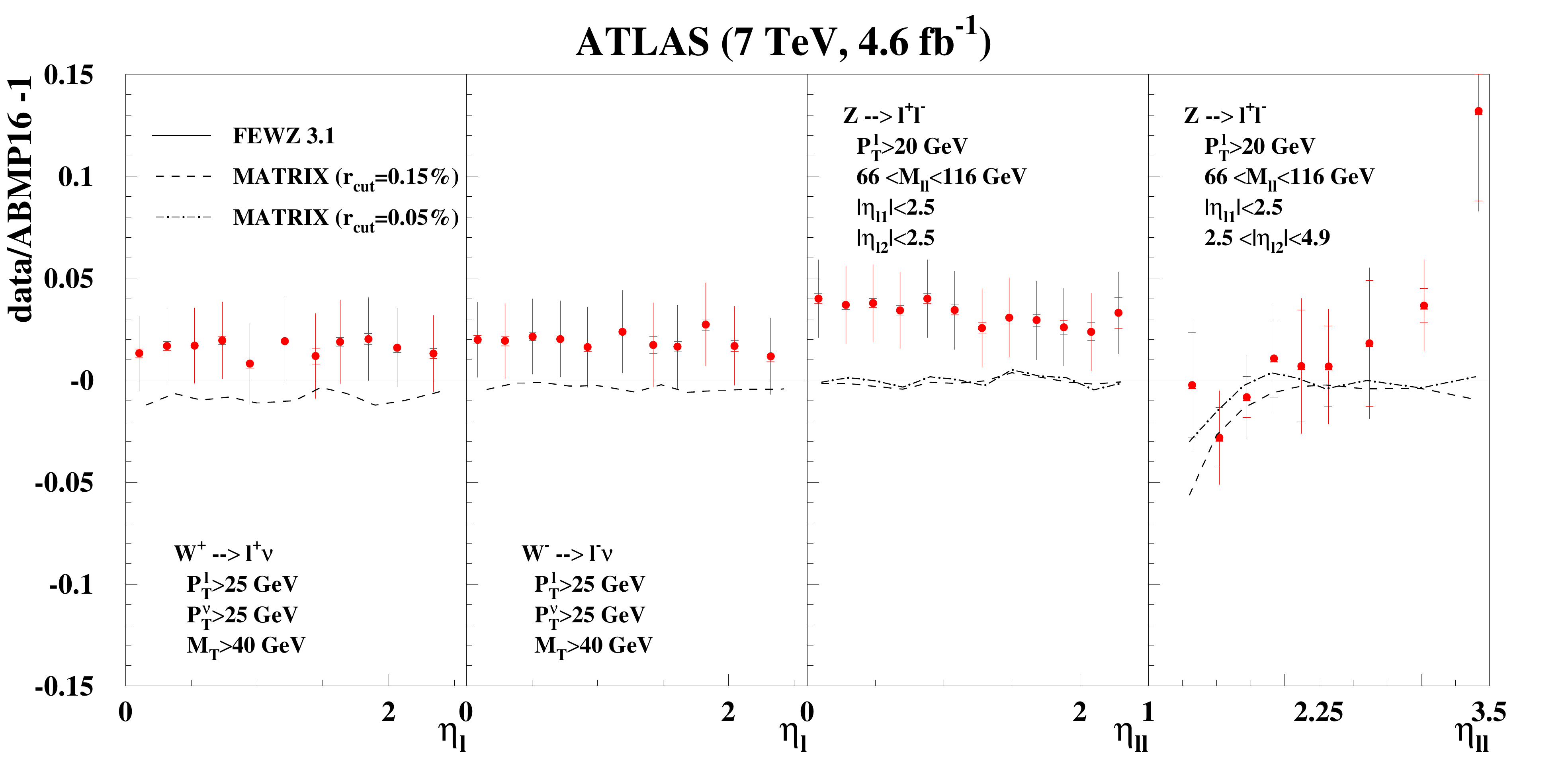}
\caption{\small
  \label{fig:matrix-atlas7data}
  Same as Fig.~\ref{fig:dynnlo-atlas7data} using predictions by the 
  {\tt MATRIX} code with different values for the $q_T$-slicing cut: $r^{\rm min}_{\rm cut}=0.15\%$ (dashed) and $r^{\rm min}_{\rm cut}=0.05\%$ (dashed-dotted).
}
\end{center}
\end{figure}
\begin{figure}[t!]
\begin{center}
\includegraphics[width=16.5cm]{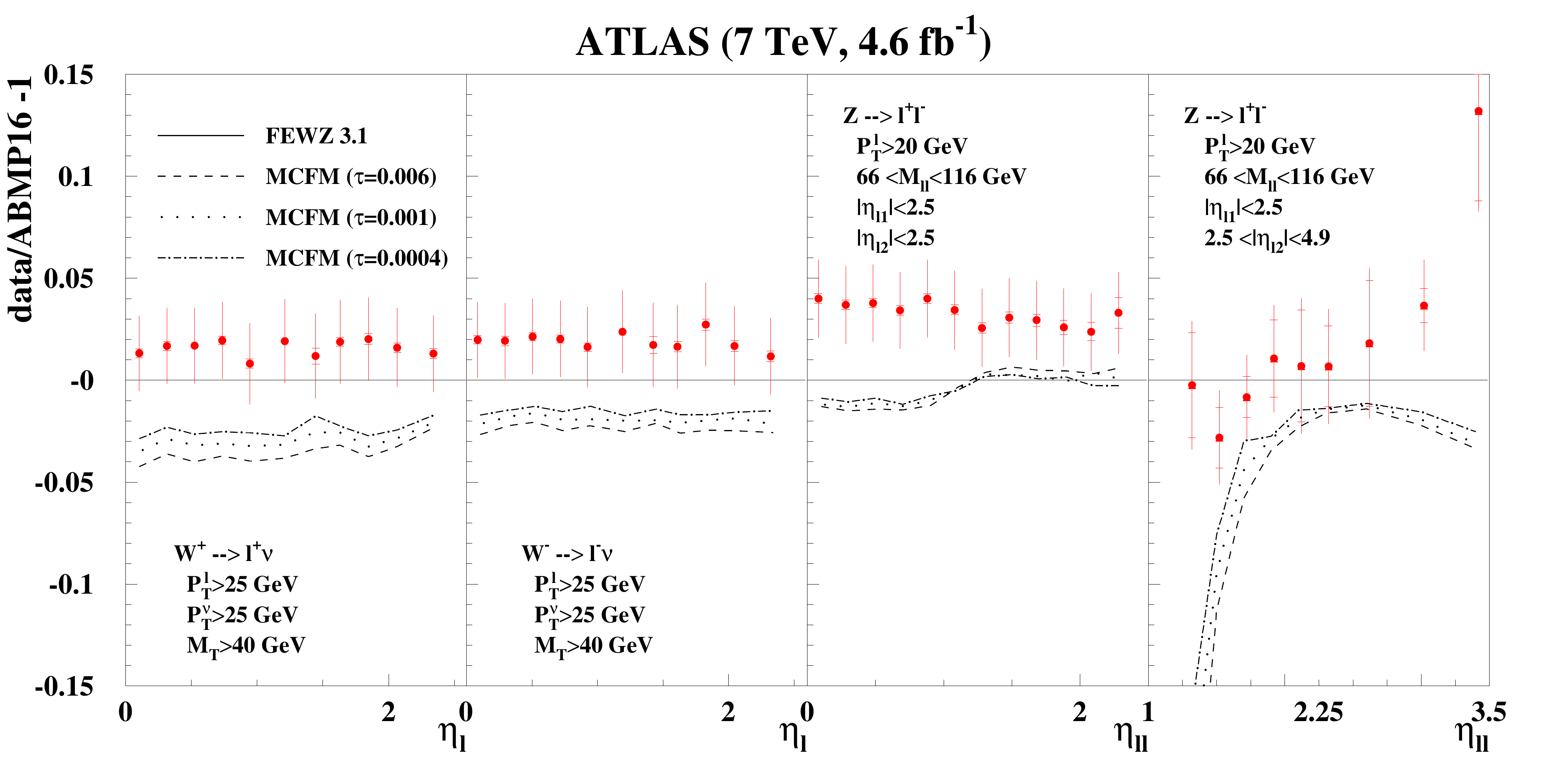}
\caption{\small
  \label{fig:mcfm-atlas7data}
  Same as Fig.~\ref{fig:dynnlo-atlas7data} using predictions by the 
  {\tt MCFM} code and with different values for the jettiness slicing parameter: 
  $\tau_{\rm cut}=6\cdot 10^{-3}$ (dashed), 
  $\tau_{\rm cut}=1\cdot 10^{-3}$ (dotted), 
  and $\tau_{\rm cut}=4\cdot 10^{-4}$ (dashed-dotted).
}
\end{center}
\end{figure}

In Fig.~\ref{fig:dynnlo-atlas7data} we show the pulls of data and the
differences of NNLO predictions obtained from {\tt DYNNLO} 
normalized to the predictions computed with the {\tt FEWZ} code
for the same distributions as studied in the previous section at NLO.
The nominal relative accuracy of the numerical integration is in all cases at the
level of a few units in $10^{-4}$, thus negligible in the plots.
For the lepton-pseudo-rapidity ($\eta_l$) distribution in $W^+$-production in
Fig.~\ref{fig:dynnlo-atlas7data}, the {\tt DYNNLO} results 
are below the {\tt FEWZ} ones by up to ${\cal O}(1\%)$ 
and by a few per mill for $W^-$-production, in both cases over the whole range in $\eta_l$. 
For the di-lepton pseudo-rapidity distribution in the central $Z/\gamma^*$-production the {\tt DYNNLO} predictions
are below the {\tt FEWZ} ones by several per mill up to 
$\eta_{ll} \leq 1.5$ and tend to agree better with the {\tt FEWZ} ones for larger $\eta_{ll}$.
For forward $Z/\gamma^*$-production instead, we see the {\tt DYNNLO} results
being below the {\tt FEWZ} ones by up to ${\cal O}(1-2\%)$ 
in the entire $\eta_{ll}$ range, with significant deviations up to ${\cal O}(7\%)$ in first bins.

In Fig.~\ref{fig:matrix-atlas7data} we show the same comparison, now with the
NNLO predictions obtained with the {\tt MATRIX} code. 
Our findings regarding the deviations from the {\tt FEWZ} results 
are qualitatively similar, quantitatively slightly smaller to those from {\tt DYNNLO}, with both codes being based on the $q_T$-slicing method.
All {\tt DYNNLO} results have been obtained with the default minimum value
$r^{\rm min}_{\rm cut}=q_T^{\rm min}/M_V=0.8\%$ for the slicing cut on $q_T$. 
The {\tt MATRIX} code uses $r^{\rm min}_{\rm cut}=0.15\%$ as the default and 
offers also the choice $r^{\rm min}_{\rm cut}=0.05\%$ for the $Z/\gamma^*$-production process.
In detail we find {\tt MATRIX} results 
being below the {\tt FEWZ} ones by up
to ${\cal O}(1\%)$ for the $\eta_l$-distribution in $W^+$-production, and by a few per
mill for $W^-$-production. 
For the $\eta_{ll}$-distribution in $Z/\gamma^*$-production at central
rapidities we see the {\tt MATRIX} numbers with the default value
$r^{\rm min}_{\rm cut}=0.15\%$ below the {\tt FEWZ} ones by a few per mill 
for $\eta_{ll} \leq 1.0$, above the {\tt FEWZ} ones in the bins around 
$\eta_{ll} \simeq 1.5$ and in agreement with {\tt FEWZ} for larger rapidities.
On the other hand, the {\tt MATRIX} results with $r^{\rm min}_{\rm cut}=0.15\%$ 
for forward $Z/\gamma^*$-production are below the {\tt FEWZ} ones by up to ${\cal O}(1-2\%)$ 
in the entire $\eta_{ll}$ range and show a significant deviation of ${\cal O}(5\%)$ in the first bin.
In order to deal with the residual dependence on the slicing cut in $q_T$ 
{\tt MATRIX} extrapolates the total rates for $r^{\rm min}_{\rm cut} \to 0$
and suggests a uniform rescaling of each bin by the ratio 
$\sigma^{\rm extrapolated}_{\rm NNLO}/\sigma^{r_{\rm cut}}_{\rm NNLO}$, see also Sec.~\ref{sec:powercorrections}.
In Fig.~\ref{fig:matrix-atlas7data} this rescaling has not been applied.
If done, it would lead to upward shifts of the central values obtained with {\tt MATRIX} 
by $5 \pm 2$ \permille\ for $W^+$- and by $4 \pm 2$ \permille\ for $W^-$-production.
Central $Z$-boson predictions would move upwards by $2 \pm 1$ \permille\ 
and the ones for forward $Z$-bosons by $7 \pm 3$ \permille.
The uncertainty in those rescaling factors comes from the extrapolation uncertainty
in $\sigma^{\rm extrapolated}_{\rm NNLO}$. Such shifts decrease, but do not eradicate the differences. 
The {\tt MATRIX} results with the smaller value $r^{\rm min}_{\rm cut}=0.05\%$ lead to better agreement with the {\tt FEWZ} results, 
i.e., there are systematic upward shifts in  Fig.~\ref{fig:matrix-atlas7data}. 
In detail, these amount to a few per mill for $\eta_{ll} \leq 1.0$ for $Z/\gamma^*$-production at central rapidities and up to a few per cent for forward $Z/\gamma^*$-production in the bins with $\eta_{ll} \lesssim 2.0$. 
The computational demands for these {\tt MATRIX} runs were huge.
\footnote{The required CPU times 
    for the {\tt MATRIX} runs with $r^{\rm min}_{\rm cut}=0.05\%$ 
were roughly 200.000 hrs for central and approximately 350.000 hrs for forward $Z$-boson production.} 
The suggested rescaling factor $\sigma^{\rm extrapolated}_{\rm
  NNLO}/\sigma^{r_{\rm cut}}_{\rm NNLO}$ turns out to be unity within the numerical accuracy of our computation for central  $Z/\gamma^*$-production. Predictions for forward $Z$-bosons would be shifted upwards uniformly by $3 \pm 2$ \permille\ and the observed differences, especially in the first $\eta_{ll}$ bins, would still persist. 

\begin{figure}[t!]
\begin{center}
\includegraphics[width=16.5cm]{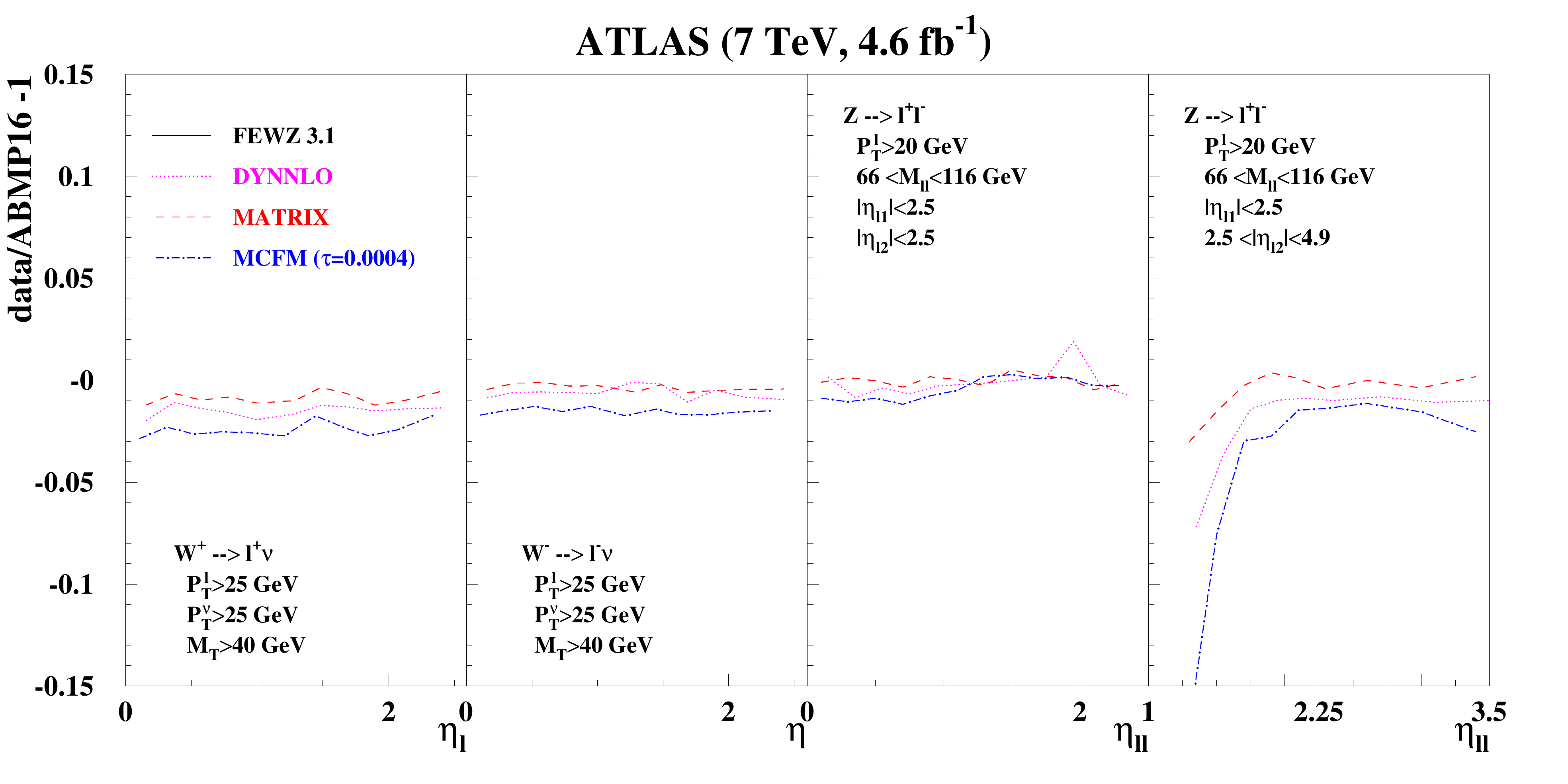}
\caption{\small
  \label{fig:allNNLO}
  Compilation of the NNLO theory predictions of Figs.~\ref{fig:dynnlo-atlas7data}--\ref{fig:mcfm-atlas7data}. 
  Only the results with the smallest slicing cuts are plotted: 
  {\tt MATRIX} with $r_{\rm cut}=0.15\%$ for $W^\pm \to l^\pm \nu$ 
  and $r_{\rm cut}=0.05\%$ for $Z\to l^+l^-$ production; 
  {\tt MCFM} with $\tau_{\rm cut}=4\cdot 10^{-4}$.}
\end{center}
\end{figure}

Finally, in Fig.~\ref{fig:mcfm-atlas7data} we repeat the benchmark study with
NNLO predictions obtained with the {\tt MCFM} code, in which case the 
numerical integration accuracy is typically ${\cal O}(1 \permille)$ and negligible in the plots.
We do find substantial deviations of the {\tt MCFM} results at NNLO, being below the {\tt FEWZ} ones 
for all distributions considered. 
Differences amount to ${\cal O}(3\%)$ for the $\eta_l$-distribution in
$W^+$-production and up to ${\cal O}(2\%)$ for $W^-$-production, respectively.
For central $Z/\gamma^*$-production the $\eta_{ll}$-distribution is also 
${\cal O}(2\%)$ below the {\tt FEWZ} results for $\eta_{ll} \leq 1.5$ 
and up to ${\cal O}(2-3\%)$ for forward $Z/\gamma^*$-production. 
In the first bins of the latter the deviations grow up to ${\cal O}(20\%)$.
As discussed, {\tt MCFM} uses $N$-jettiness subtraction and allows for 
different $\tau_{\rm cut}$ choices for the jettiness slicing parameter. 
We use the default value, $\tau_{\rm cut}=6\cdot 10^{-3}$ and two smaller ones, 
$\tau_{\rm cut}=1\cdot 10^{-3}$ and $\tau_{\rm cut}=4\cdot 10^{-4}$, the
limitation being here the goal to reach an integration accuracy  
of a few units in $10^{-4}$ in reasonable time 
\footnote{The required CPU times for the {\tt MCFM} runs with $\tau_{\rm cut}=4\cdot 10^{-4}$ 
were roughly 180.000 hrs for $W^\pm$-boson, 160.000 hrs for central and approximately 50.000 hrs for forward $Z$-boson production.} 
with given computational resources.
The decreasing values of $\tau_{\rm cut}$ display the expected trend clearly
in Fig.~\ref{fig:mcfm-atlas7data}, namely, the smaller the choice of $\tau_{\rm cut}$,
the closer the {\tt MCFM} result to that by {\tt FEWZ}. Nevertheless, the differences remain.
In order to compare those differences easier, we collect the best prediction for each code at NNLO in a single figure in Fig.~\ref{fig:allNNLO}.

Given the level of agreement among the predictions at NLO accuracy, 
the deviations observed in Figs.~\ref{fig:dynnlo-atlas7data}--\ref{fig:mcfm-atlas7data}
need to be put into perspective by looking at the size of the
pure NNLO corrections alone, which we define bin-by-bin 
through the deviation of the NNLO K-factor from one, $\delta_{\rm NNLO} = (\sigma_{\rm NNLO}/\sigma_{\rm NLO}-1)$.
Typically pure NNLO corrections $\delta_{\rm NNLO}$ are rather small, and we illustrate those only in the case of largest corrections.
For $W^+$-production $\delta_{\rm NNLO}$ amounts to a few per mill for 
$\eta_{l} \lesssim 1$ and grows to ${\cal O}(1-2\%)$ for larger rapidities
$\eta_{l} \gtrsim 1$, while instead for $W^-$-production 
$\delta_{\rm NNLO}$ is of the size ${\cal O}(1\%)$
for $\eta_{l} \lesssim 1$ and increases to a few per cent for larger rapidities.
For the central $Z/\gamma^*$-production the NNLO corrections 
$\delta_{\rm NNLO}$ are only a few per mill for $\eta_{ll} \lesssim 1.5$ and
grow to ${\cal O}(2-3\%)$ for larger di-lepton rapidities.
Thus, the observed differences between considered codes are actually similar in size to that of the pure NNLO corrections, even exceeding them at times.
The case of forward $Z/\gamma^*$-production features larger higher order
corrections and will be discussed in detail next.
The comparable size of the NNLO corrections and differences among the predictions signal that the numerical precision of the studied computer programs does not match the formal accuracy of predictions at NNLO.

\begin{figure}[t!]
\begin{center}
\includegraphics[width=8.15cm]{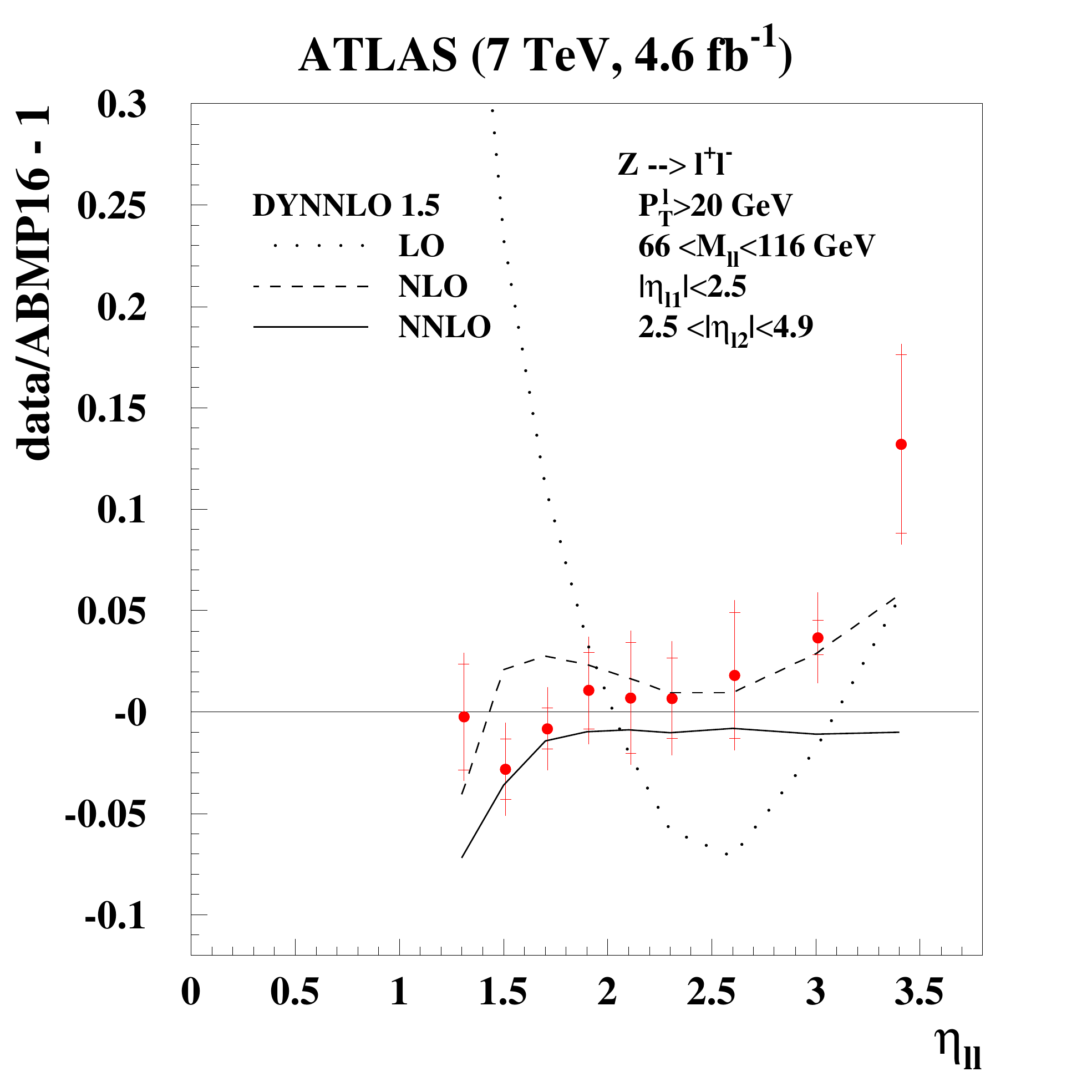}
\includegraphics[width=8.15cm]{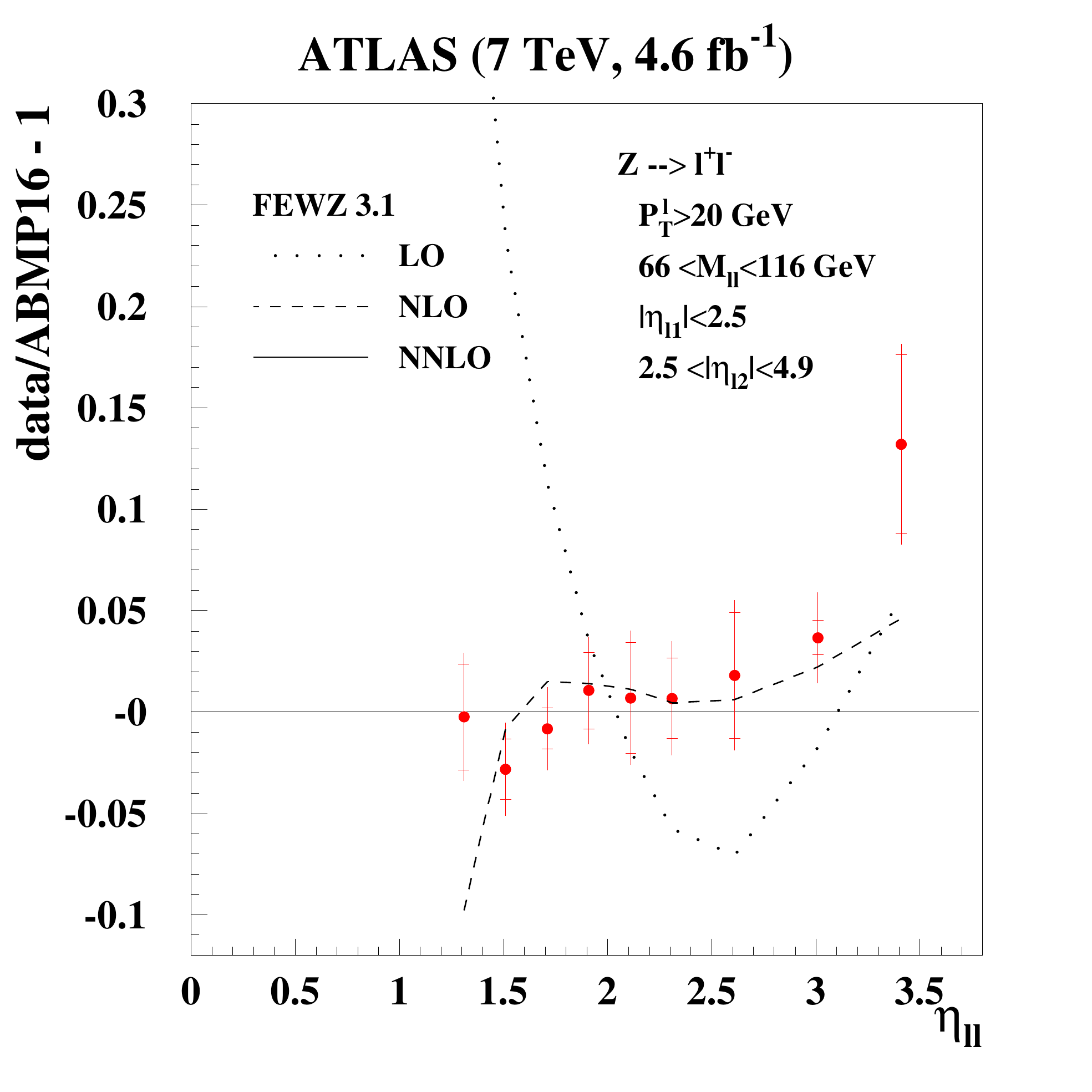}
\caption{\small
  \label{fig:dynnlozf}
  The pulls for the ATLAS data for $pp \to Z/\gamma^*+X \to l^+l^-  + X$ production 
  at forward rapidities measured at $\sqrt{s}=7$~TeV~\cite{Aaboud:2016btc},
  normalized to the ABMP16 predictions at 
  NNLO obtained with {\tt FEWZ} (version 3.1) compared to predictions 
  by the {\tt DYNNLO} (left) and the {\tt FEWZ} codes (right).
  Shown are the LO (dotted), NLO (dashed) and NNLO (solid) predictions for
  each code.}
\end{center}
\end{figure}

\begin{figure}[t!]
\begin{center}
\includegraphics[width=8.15cm]{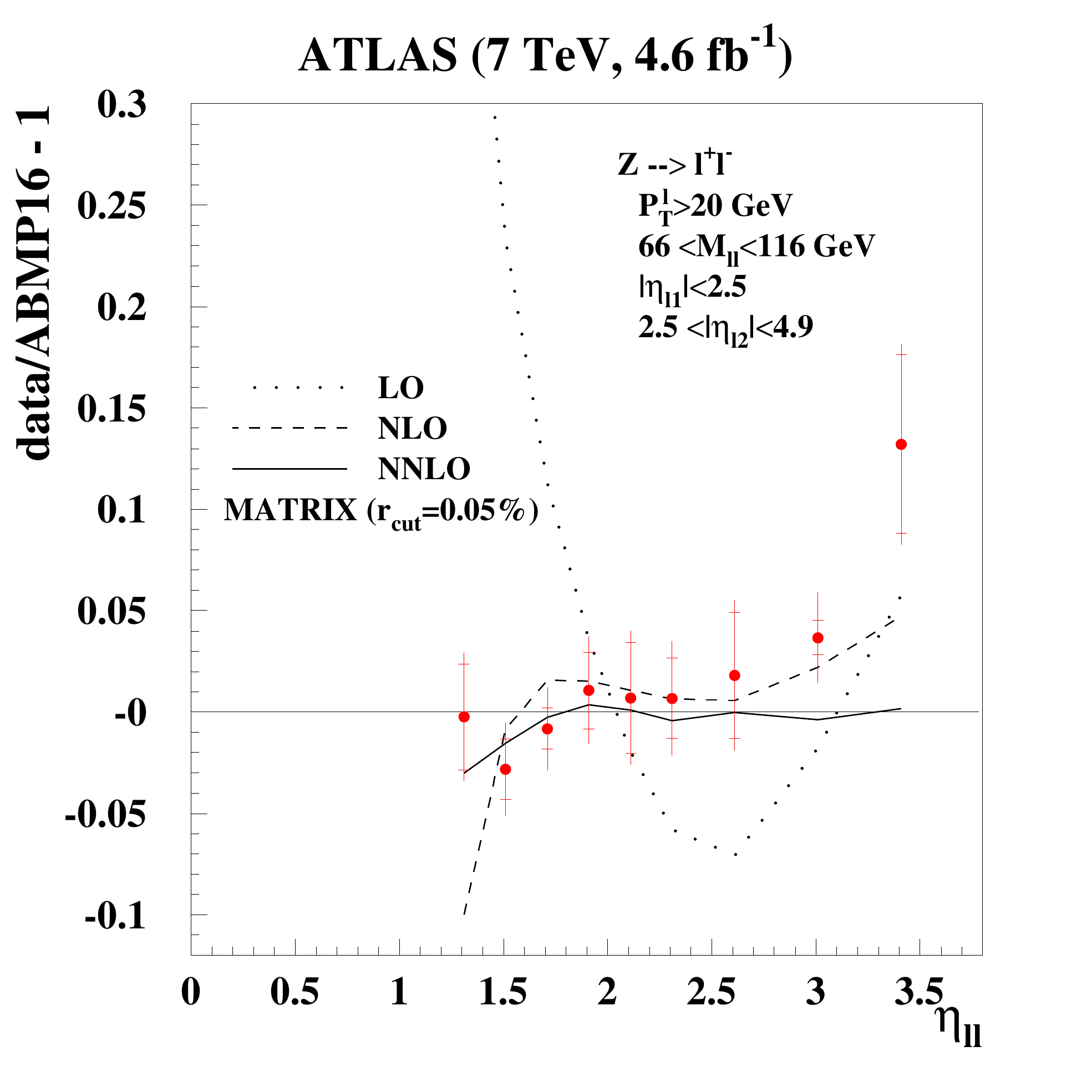}
\includegraphics[width=8.15cm]{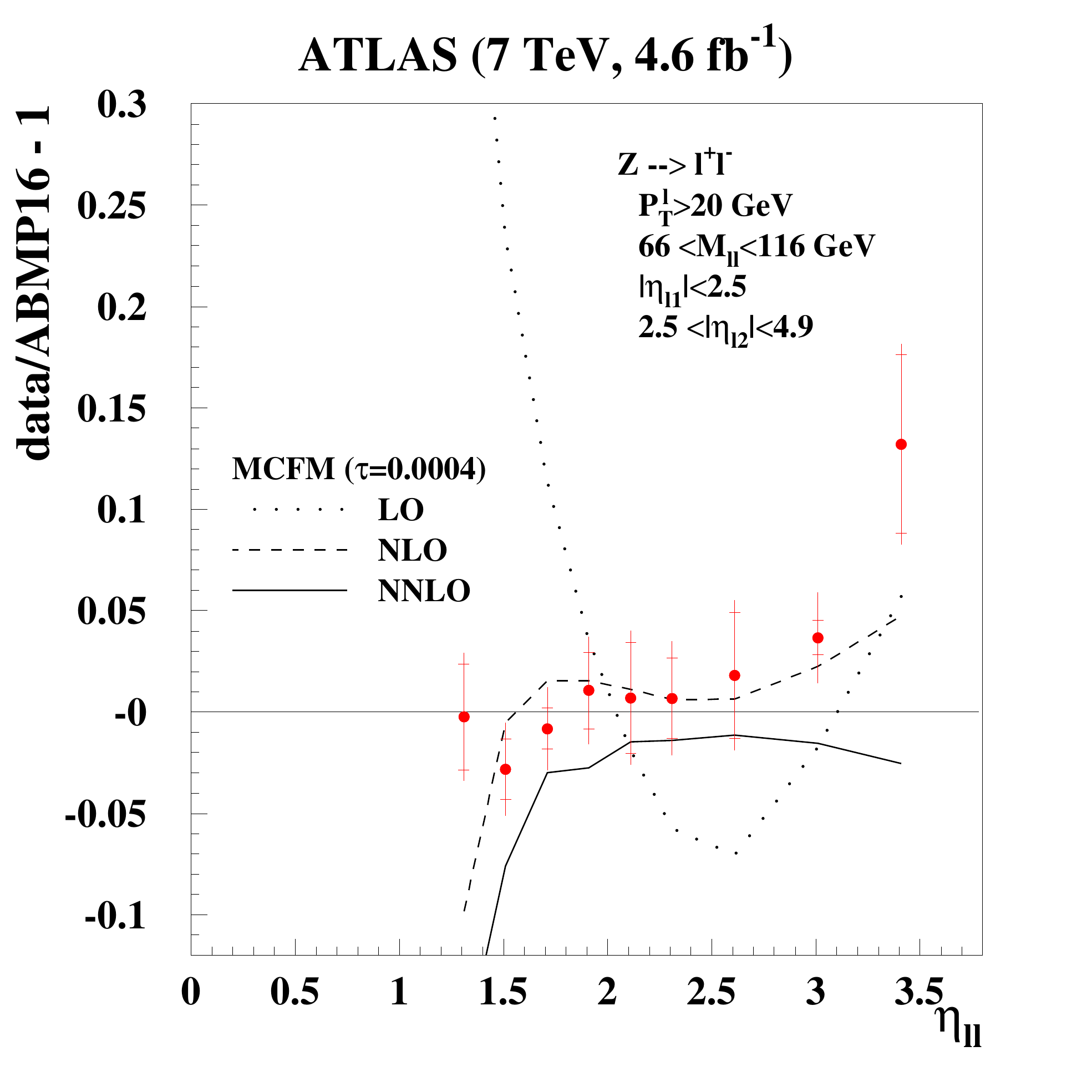}
\caption{\small
  \label{fig:mcfmzf}
  Same as Fig.~\ref{fig:dynnlozf} using predictions 
  by the {\tt MATRIX} (left) and the {\tt MCFM} codes (right).
}
\end{center}
\end{figure}

In Fig.~\ref{fig:dynnlozf} we focus on the $\eta_{ll}$-distribution for the forward $Z/\gamma^*$-production 
obtained by ATLAS~\cite{Aaboud:2016btc}. 
The particular fiducial cuts on the decay leptons for these data lead to
sizeable QCD corrections at higher orders, which we illustrate in Fig.~\ref{fig:dynnlozf}, 
where we display at LO, NLO and NNLO accuracies obtained with {\tt DYNNLO} (left) and with {\tt FEWZ} (right).
The same comparison is performed in Fig.~\ref{fig:mcfmzf} for 
the results of the {\tt MATRIX} and the {\tt MCFM} codes, where 
we display $\sigma_{\rm NNLO}$ with the smallest slicing cuts, 
$r^{\rm min}_{\rm cut}=0.05\%$ and $\tau_{\rm cut}=4\cdot 10^{-4}$.
As already remarked above, we use ABMP16 PDFs at NNLO in all cases,
independent of the perturbative order.
Figs.~\ref{fig:dynnlozf} and~\ref{fig:mcfmzf} clearly illustrate the significant corrections up to 
${\cal O}(50\%)$ in first bins, when increasing the perturbative order from LO to NLO, 
while the change from NLO to the NNLO QCD predictions still amounts to 
corrections of ${\cal O}(5-10\%)$ in some $\eta_{ll}$ bins.
The LO results in Figs.~\ref{fig:dynnlozf} and~\ref{fig:mcfmzf} are all in
perfect agreement and the deviations in the NLO predictions by {\tt DYNNLO} have already been discussed above.

The observed pattern of the higher order corrections for the predictions with {\tt FEWZ}
in Fig.~\ref{fig:dynnlozf} (right) and with {\tt MATRIX} in Fig.~\ref{fig:mcfmzf} (left) 
is very similar. 
The overall offset of the pulls for the {\tt MATRIX} results with $r^{\rm min}_{\rm cut}=0.05\%$ compared to
the {\tt FEWZ} ones is small in the entire $\eta_{ll}$ range except for the first $\eta_{ll}$ bins and 
originates from the different NNLO cross sections as illustrated in 
Fig.~\ref{fig:matrix-atlas7data}.
In contrast, the cross sections $\sigma_{\rm LO}$, $\sigma_{\rm NLO}$ and $\sigma_{\rm NNLO}$ 
from {\tt DYNNLO} in Fig.~\ref{fig:dynnlozf} (left) 
and from {\tt MCFM} in Fig.~\ref{fig:mcfmzf} (right) show a different trend. 
The pulls for the ATLAS data follow rather closely the respective 
NLO predictions across the entire range in rapidities. 
The NNLO predictions from those codes do undershoot the data by several per cent, 
which causes the significant deviations displayed in 
Figs.~\ref{fig:dynnlo-atlas7data} and~\ref{fig:mcfm-atlas7data}.

Next we continue the benchmark studies with D{\O} data on the electron charge asymmetry distribution $A_e$, 
which has been obtained as a function of the electron pseudo-rapidity from 
$W^\pm$-boson production at $\sqrt{s}=1.96$~TeV at the Tevatron~\cite{D0:2014kma}.
This observable is also subject to larger higher order corrections so that we
illustrate again the size of the LO, NLO and NNLO predictions obtained, 
as before, in all cases with the NNLO ABMP16 PDFs and $\alpha_s^{(n_f=5)}(M_Z) = 0.1147$ 
and we plot the difference to the NNLO predictions computed with the {\tt FEWZ} code.
The D{\O} data had already been included in the fit 
of the ABMP16 PDFs and a good description of those data
in the fit had been reached.
\begin{figure}[t!]
\begin{center}
\includegraphics[width=8.1cm]{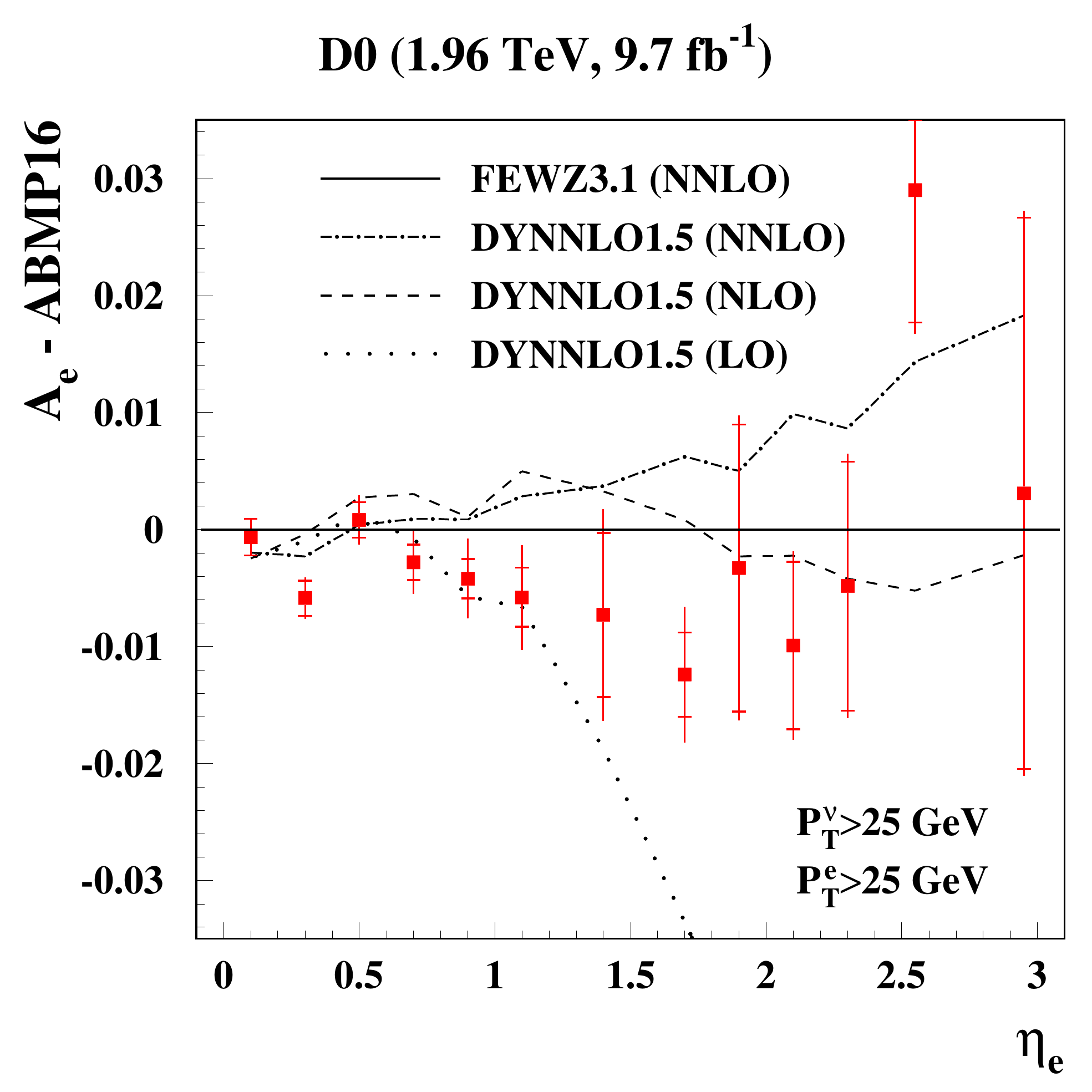}
\includegraphics[width=8.1cm]{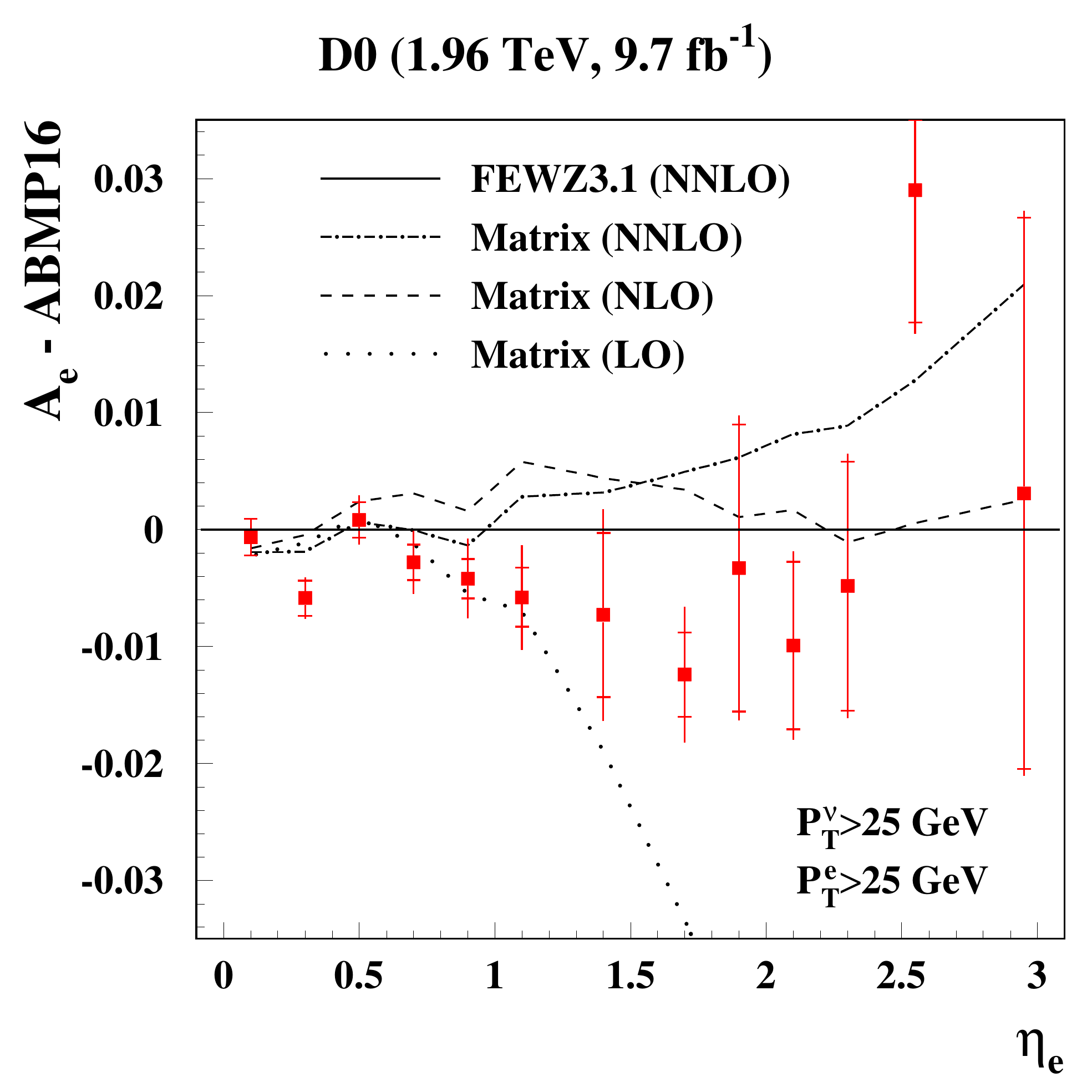}
\caption{\small
  \label{fig:dynnlo-d0}
  The D{\O} data on the electron charge
  asymmetry distribution $A_e$ in $W^{\pm}$-boson production at $\sqrt{s}=1.96$~TeV with the 
  statistical (inner bar) and the total uncertainties, including the 
  systematic ones. 
  Shown is the difference of $A_e$ to the ABMP16 central predictions at NNLO
  obtained with {\tt FEWZ}. 
  The symmetric $p_T^{e,\nu}$-cuts of the decay leptons are indicated in the figure. 
  The LO (dotted), NLO (dashed-dotted) and NNLO (dashed) predictions by 
  the {\tt DYNNLO} code (left) and by the {\tt MATRIX} code (right)
  are displayed for comparison.
}
\end{center}
\end{figure}

In Fig.~\ref{fig:dynnlo-d0} we plot in addition to the D{\O} data on $A_e$ 
the LO, NLO and NNLO predictions by the {\tt DYNNLO} code (left) and the {\tt MATRIX} code (right), 
keeping again a relative numerical integration accuracy of a few units in $10^{-4}$ for the respective $W^\pm$-boson cross sections.
The LO and NLO curves illustrate the sizable higher order corrections and 
those predictions agree among these codes.
With the given accuracy of the D{\O} data on $A_e$, also the NNLO corrections
are relevant, but we see both, the {\tt DYNNLO} and the {\tt MATRIX} results (here with $r^{\rm min}_{\rm cut}=0.15\%$) being mostly above the {\tt FEWZ} numbers.
The deviations increase with increasing electron pseudo-rapidity $\eta_e$ and 
become significant for $\eta_e \gtrsim 1.0$, where the size of the 
difference exceeds the size of the pure NNLO corrections.
For the asymmetry $A_e$ any overall rescaling of cross sections as suggested
for the {\tt MATRIX} code and described in Sec.~\ref{sec:powercorrections} to account for $r^{\rm min}_{\rm cut}$ dependence has no effect.
\begin{figure}[t!]
\begin{center}
\includegraphics[width=8.1cm]{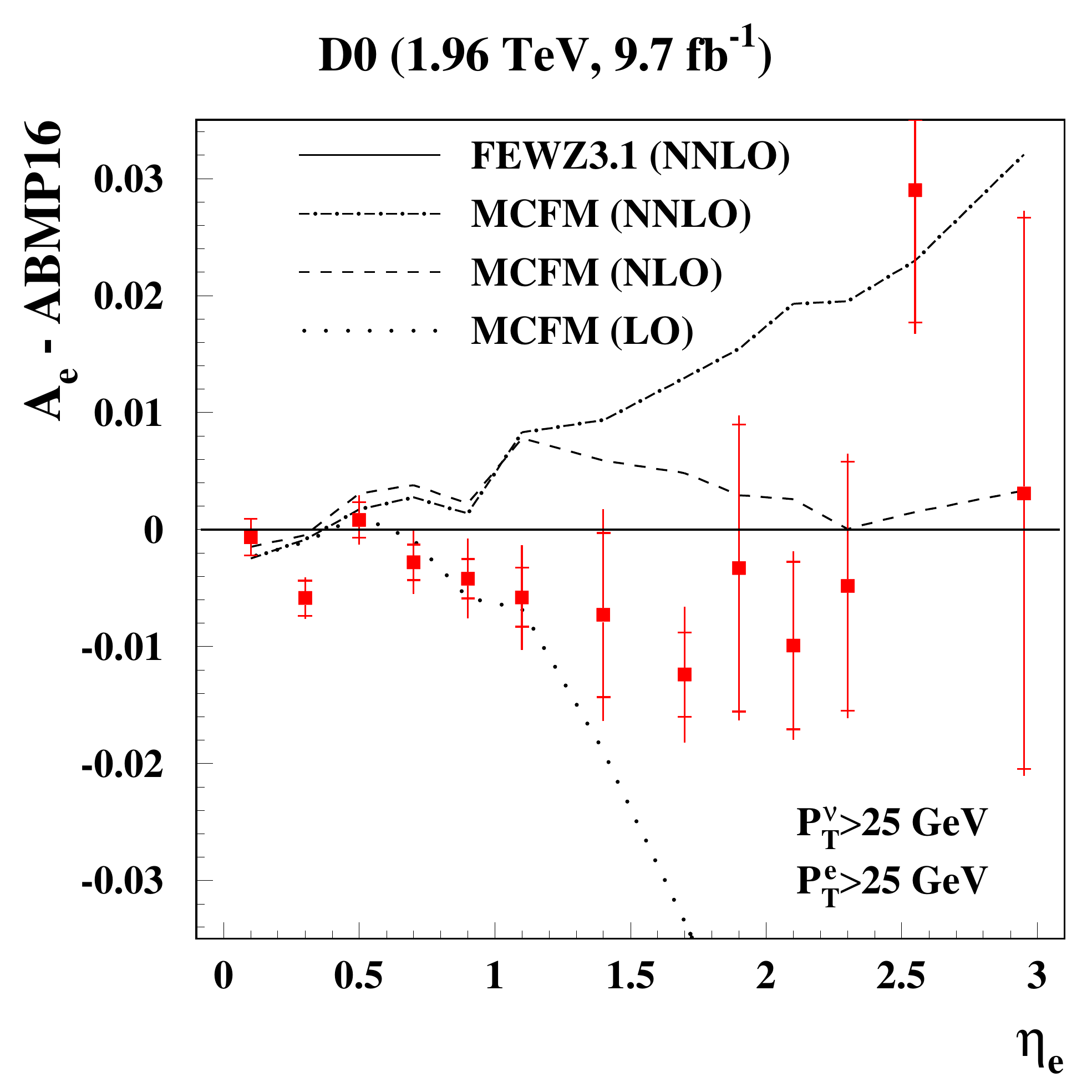}
\caption{\small
  \label{fig:mcfm-d0}
  Same as Fig.~\ref{fig:dynnlo-d0} using predictions by the {\tt MCFM} code.
}
\end{center}
\end{figure}
\begin{figure}[t!]
\begin{center}
\includegraphics[width=8.1cm]{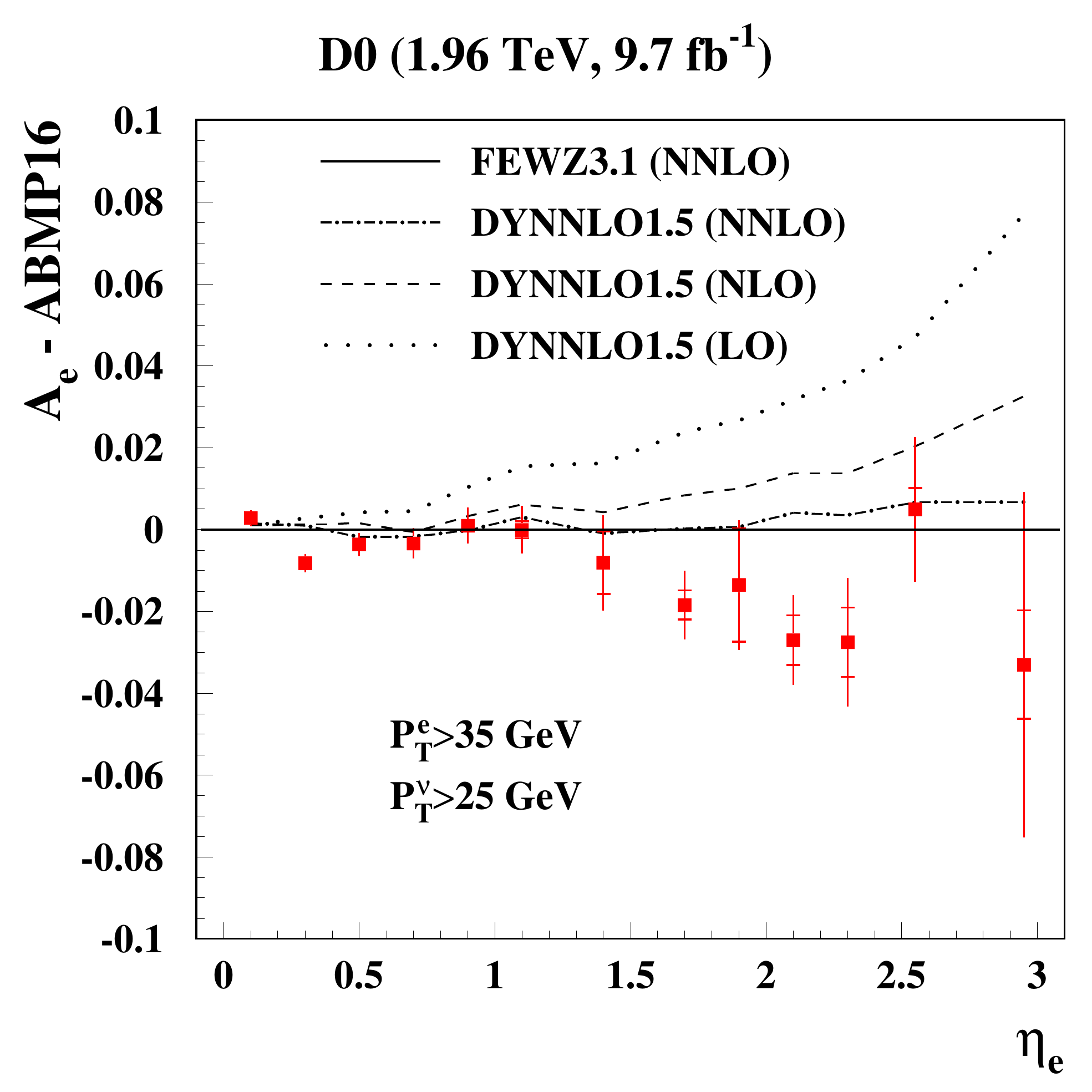}
\includegraphics[width=8.1cm]{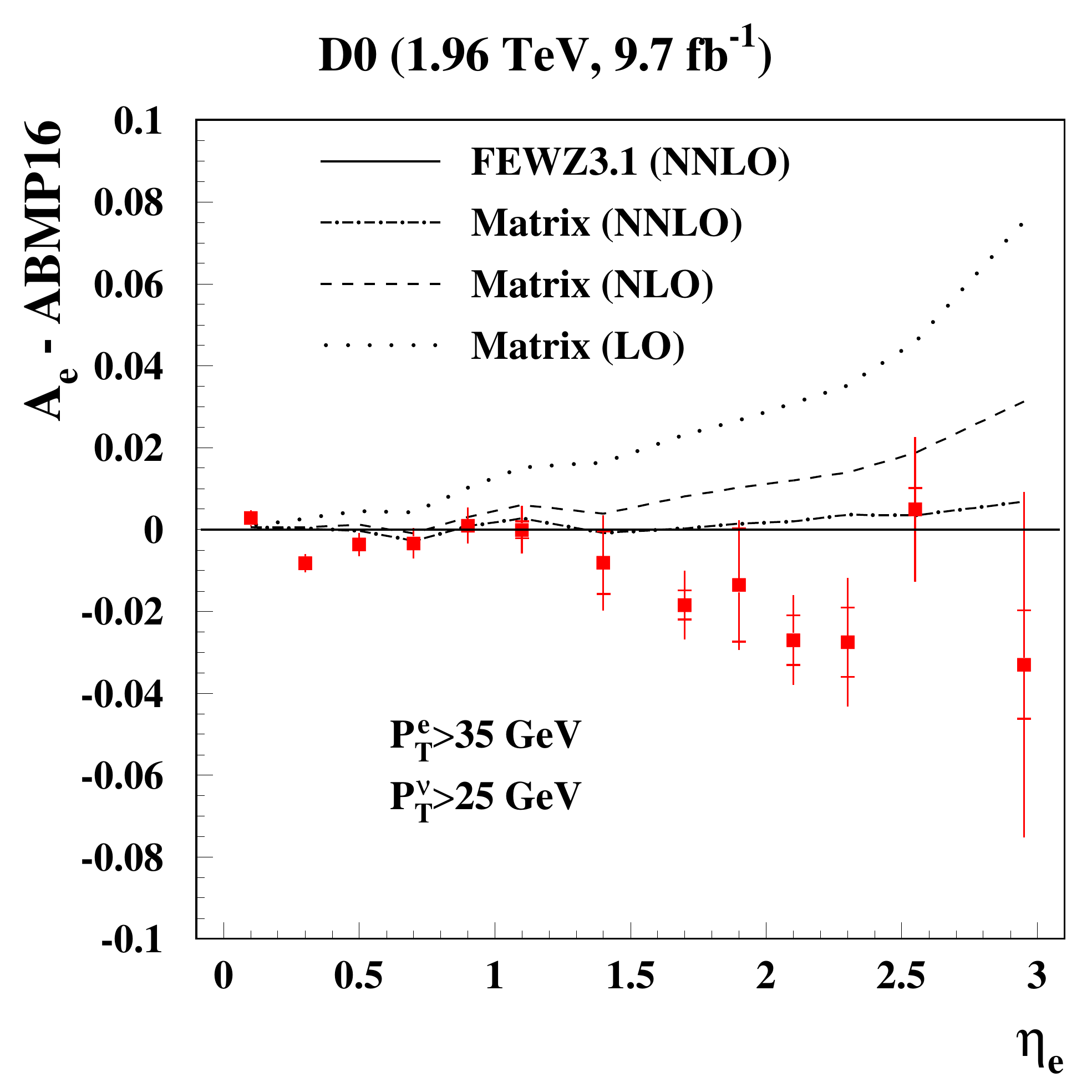}
\caption{\small
  \label{fig:dynnlo-d0-staggered}
  Same as Fig.~\ref{fig:dynnlo-d0}, now with staggered cuts 
  $p_T^{e} > p_T^{\nu}$ on the decay leptons as indicated in the figure for 
  the {\tt DYNNLO} code (left) and the {\tt MATRIX} code (right).
}
\end{center}
\end{figure}
\begin{figure}[t!]
\begin{center}
\includegraphics[width=8.1cm]{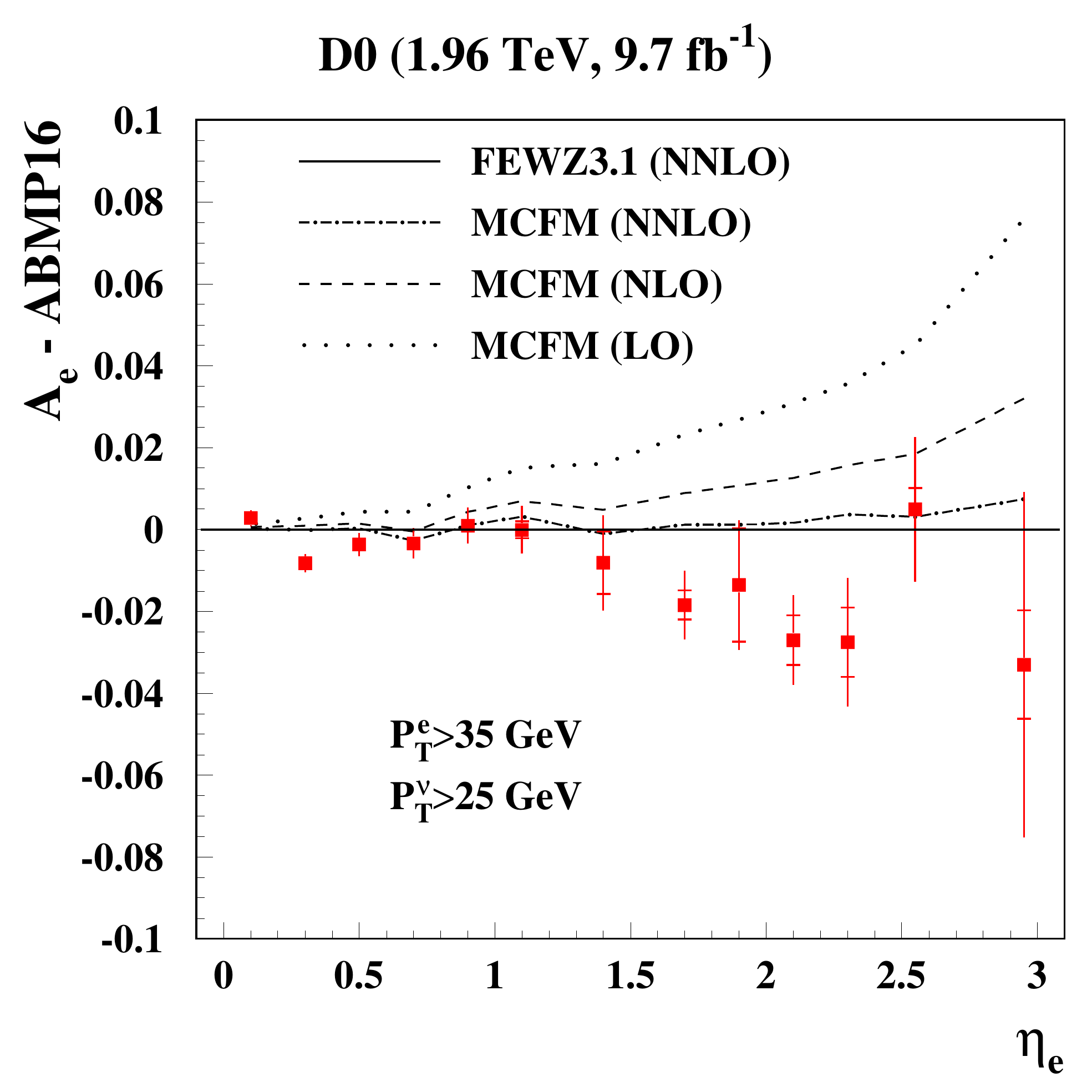}
\caption{\small
  \label{fig:mcfm-d0-staggered}
  Same as Fig.~\ref{fig:dynnlo-d0-staggered} using predictions by the {\tt MCFM} code.
}
\end{center}
\end{figure}
In Fig.~\ref{fig:mcfm-d0} we show the same study, now comparing to the results obtained with the {\tt MCFM} code. 
The NNLO {\tt MCFM} result has been computed with the default $\tau_{\rm cut}$ value, $\tau_{\rm cut}=6\cdot 10^{-3}$,
and the numerical integration accuracy of the individual $W^\pm$-boson cross sections is typically ${\cal O}(1 \permille)$. 
In addition, deviating from the default settings of {\tt MCFM}, 
the parameter {\tt cutoff} has been changed to $10^{-6}$ from $10^{-9}$, which
is its default value~\footnote{
Execution of {\tt MCFM} with the command {\tt ./mcfm\_omp input.ini -extra\%cutoff=1d-6}.}.
The parameter {\tt cutoff} provides the minimum value on any dimensionless variables, for instance, any invariant mass 
squared $s_{ij}$ of any pair of partons scaled by their energies, such that $s_{ij}/(E_i E_j)$ can never be less than {\tt cutoff}.
Cross sections $\sigma_{\rm NNLO}$ for $W^\pm$-boson production in the D{\O}
kinematics computed with {\tt cutoff}~$=10^{-9}$ showed severe numerical instabilities.
For the agreement of the NNLO {\tt MCFM} results with {\tt FEWZ}, we find a
similar pattern of increasing deviations with increasing electron
pseudo-rapidity $\eta_e$, which become significant for $\eta_e \gtrsim 1.0$. 
The two choices of smaller $\tau_{\rm cut}$ values in {\tt MCFM}, $1\cdot 10^{-3}$ and $4\cdot 10^{-4}$, lead to the same NNLO predictions for $A_e$,  within the numerical uncertainites.

Finally, Figs.~\ref{fig:dynnlo-d0-staggered} and \ref{fig:mcfm-d0-staggered}
show the D{\O} data and the predictions of {\tt DYNNLO}, {\tt MATRIX} and {\tt MCFM} relative to {\tt  FEWZ} 
for the electron charge asymmetry distribution $A_e$ in the case of staggered
cuts with $p_T^{e} \geq 35$~GeV and $p_T^{\nu} \geq 25$~GeV.
The NNLO results for {\tt MCFM} were obtained with {\tt cutoff}~$=10^{-6}$ as
in the case of symmetric cuts to avoid numerical instabilities.
Interestingly, the very good agreement among all codes already observed at NLO
in Fig.~\ref{fig:d0-asy-nlo}, is found now in all cases also at NNLO.
This is in complete contrast to the case of symmetric cuts in Figs.~\ref{fig:dynnlo-d0} and \ref{fig:mcfm-d0}.

\section{Power corrections}
\label{sec:powercorrections}

Our comparison in the previous section was based on computer codes implementing different approaches to the regularization of double real singular emissions. The global slicing methods neglect power corrections, hence one may assume that those are at least partly responsible for the observed differences, which we explore in this section.
There are two sources of power corrections, namely those intrinsic to the
$q_T$ and ${\cal T}_N$ factorization on the one hand, which have been studied
extensively in the literature before, 
see e.g.~\cite{Moult:2017jsg,Boughezal:2018mvf,Ebert:2018gsn,Cieri:2019tfv}, 
and fiducial power corrections on the other, which have been studied in detail more recently~\cite{Ebert:2019zkb}.
The latter ones arise whenever one considers fiducial cuts or leptonic observables and formally dominate.

\subsection{Review of global slicing methods}

We briefly review global slicing methods applied to the Drell-Yan process and 
with emphasis on the power counting in the slicing parameter $\tau$, see e.g., \cite{Boughezal:2018mvf,Ebert:2019zkb}. 
Specifically, we focus on the $q_T$ and $N$-jettiness subtractions~\cite{Catani:2007vq,Boughezal:2015dva,Gaunt:2015pea} 
for the hadro-production of a gauge boson $V$, which decays 
to a leptonic final state $L$ with the following kinematics
\begin{eqnarray}
  \label{eq:V}
  a(p_a)+b(p_b) &\to V(q)+X(k_i) &\to L(q)+X(k_i)
  \, ,
\end{eqnarray}
where $a, b$ are the initial hadrons with momenta $p_{a,b}$, 
$q$ is the gauge boson momentum 
and $X(k_i)$ denotes hadronic final states with momenta $k_i$.
In the laboratory frame, spanned by light-like vectors $n^\mu=(1,0,0,1)$ and ${\bar n}^\mu=(1,0,0,-1)$,
the initial hadron momenta $p_{a,b}$ can be parametrized in Born kinematics 
in terms of $Q=\sqrt{q^2}$ and the rapidity of the gauge boson $Y$ as 
\begin{eqnarray}
\label{eq:pa-pb-para}
p_a^\mu\,=\, \frac{Q}{2}\,\exp(+Y)\,n^\mu\, , \qquad\qquad
p_b^\mu\,=\, \frac{Q}{2}\,\exp(-Y)\,{\bar n}^\mu
 \, .
\end{eqnarray}
The slicing parameter $\tau$ for $q_T$-subtraction is defined by
\begin{eqnarray}
\label{eq:tau-qt}
  \tau &=\, q_T^2/Q^2 &=\, \left(\sum_i \vec{k}_{T,i}\right)^2/Q^2 
  \, ,
\end{eqnarray}
in terms of the transverse momenta $\vec{k}_{T,i}$ of the hadronic real emission radiation.
Likewise, one has 
\begin{eqnarray}
\label{eq:tau-0jet}
  \tau &=\, {\cal T}_0/Q &=\, \sum_i {\rm min} \left\{ 2 p_a \cdot k_i, 2 p_b \cdot k_i \right\}/Q^2 
  \, ,
\end{eqnarray}
for the leptonic $0$-jettiness ${\cal T}_0$ (see \cite{Boughezal:2018mvf,Ebert:2019zkb} for other variants of $0$-jettiness definitions).
The leptonic ${\cal T}_0$ is preferred due to smaller intrinsic power corrections, cf. \cite{Moult:2016fqy}
and used in {\tt MCFM}~\cite{Campbell:2019dru}.
The definitions of $\tau$ in eqs.~(\ref{eq:tau-qt}) and (\ref{eq:tau-0jet}) 
vanish at Born level and resolve additional radiation in an infrared-safe manner, 
so that the phase space integration for the cross section can be written as 
\begin{eqnarray}
  \label{eq:crs-tau}  
\sigma &=& \int d\tau\, \frac{d\sigma}{d\tau} 
\,=\,
\int^{\tau_{\rm cut}}\!\! d\tau\, \frac{d\sigma}{d\tau} 
+ \int_{\tau_{\rm cut}}\! d\tau\, \frac{d\sigma}{d\tau} 
\,=\,
\sigma(\tau_{\rm cut}) 
+ \int_{\tau_{\rm cut}}\! d\tau\, \frac{d\sigma}{d\tau} 
\, ,
\end{eqnarray}
where $\tau_{\rm cut}$ is the cut for the slicing of the phase space. 
The dependence of $d\sigma/d\tau$ on $\tau$ can be predicted from the universal factorization of QCD in soft and collinear limits. 
It has the structure 
\begin{eqnarray}
  \label{eq:sigma-tau}  
\frac{d\sigma}{d\tau} &\sim& \delta(\tau) + \sum_i \left[\frac{\ln^i \tau}{\tau}\right]_+ + \sum_j \tau^{p-1} \ln^j \tau + {\cal O}(\tau^{p})
\, ,
\end{eqnarray}
where the $+$-distributions are the well-known leading threshold logarithms and the terms 
proportional to $\tau^{p-1}$ with $p>0$ are integrable and denote power corrections in the soft and collinear limit. 
From the analytical integration one obtains for $\sigma(\tau_{\rm cut})$ schematically
\begin{eqnarray}
  \label{eq:sigma-taucut}  
\sigma(\tau_{\rm cut}) &\sim& 1 + \sum_i \ln^{i+1} \tau_{\rm cut} + \sum_j \tau^{p}_{\rm cut} \ln^{j} \tau_{\rm cut} + {\cal O}(\tau_{\rm cut}^{p+1})
\, .
\end{eqnarray}
The crucial point to stress here is the scaling behavior of the power corrections $\tau^{p}_{\rm cut}$, i.e. the value of the exponent $p$.
For the production of a stable gauge boson $V$, $p$ takes positive integer values, 
while the subsequent decay with cuts on the leptonic final state changes the scaling of the power corrections~\cite{Ebert:2019zkb}, 
such that $p$ rises in steps of half-integers, i.e., $p=1/2,1,3/2$ and so on. 
This will be discussed in more detail below.

The scaling of the power corrections has consequences for the particular subtraction scheme, 
which is then implemented via a global subtraction term $\sigma^{\rm sub}(\tau_{\rm cut})$ as 
\begin{eqnarray}
  \label{eq:crs-sub}  
\sigma &=& 
\sigma^{\rm sub}(\tau_{\rm cut}) 
+ \int_{\tau_{\rm cut}}\! d\tau\, \frac{d\sigma}{d\tau} 
+ \Delta \sigma^{\rm sub}(\tau_{\rm cut}) 
\, .
\end{eqnarray}
Here the term $\Delta \sigma^{\rm sub}(\tau_{\rm cut}) = \sigma(\tau_{\rm cut}) - \sigma^{\rm sub}(\tau_{\rm cut})$ 
parametrizes the residual power corrections. 
It is neglected in slicing methods, giving rise to an intrinsic error of these methods. 
If the global subtraction term $\sigma^{\rm sub}(\tau_{\rm cut})$ cancels only 
the leading soft and collinear singularities in $\sigma(\tau_{\rm cut})$ in eq.~(\ref{eq:sigma-taucut}), 
then the residual power corrections in the presence of cuts on the decay
leptons scale as $\sqrt{\tau_{\rm cut}}$. 
This implies enhanced corrections of the order $q_T/Q$ for the $q_T$
subtraction, as will be explained below, or of the order of $\sqrt{{\cal T}_0/Q}$ for the $N$-jettiness subtraction, as detailed
in~\cite{Ebert:2019zkb} with a power counting argument.

The phase space slicing codes under consideration employ different strategies
for dealing with power corrections. 
{\tt MATRIX} performs an extrapolation of $r_{\rm cut}=q_T/M_V \to 0$ 
for the total rate of the process computed with $q_T$-subtraction 
by evaluating the cross section at fixed values in the interval
$r_{\rm cut} \in [0.15,1]\%$ in steps of $0.01\%$.
It is then recommended to correct the kinematic distributions 
by rescaling uniformly with the ratio $\sigma^{\rm extrapolated}_{\rm NNLO}/\sigma^{r_{\rm cut}}_{\rm NNLO}$.
{\tt MCFM} has improved the $\tau_{\rm cut}$ dependence by implementing the leading power
corrections of~\cite{Moult:2016fqy,Ebert:2018lzn} (see also \cite{Boughezal:2018mvf}),
which are derived for the production of stable gauge bosons $V$ and scale as $\tau_{\rm cut}$, 
cf. eq.~(\ref{eq:sigma-taucut}).
In addition, {\tt MCFM} computes the cross section for an array of different $\tau_{\rm cut}$ values 
and performs an automated fitting of the $\tau_{\rm cut}$ dependence 
at NNLO with the following ansatz 
\begin{eqnarray}
  \label{eq:taucut-fitnnlo} 
  \sigma(\tau_{\rm cut})^{\rm NNLO} &=& \sigma_0 
  + c_1\, \tau_{\rm cut} \ln^3(\tau_{\rm cut}/M_V) + c_2\, \tau_{\rm cut} \ln^2(\tau_{\rm cut}/M_V) 
  + c_3\, \tau_{\rm cut}
  \, ,
\end{eqnarray}
where $c_i$ are the fit parameters and the result is then extrapolated to $\tau_{\rm cut} \to 0$.
Note, that the functional form in eq.~(\ref{eq:taucut-fitnnlo}) does not 
capture well the scaling of the leading power corrections proportional to $\sqrt{\tau_{\rm cut}}$ 
in the case of gauge boson decays~\cite{Ebert:2019zkb}.

\subsection{Fiducial cuts}

For the discussion of the fiducial cuts on the decay leptons, 
we follow the presentation in \cite{Ebert:2019zkb,Ebert:2020dfc}.
First, we need to further specify the leptonic final state $L$ in eq.~(\ref{eq:V}), which reads 
\begin{eqnarray}
  \label{eq:L}
Z/\gamma^* \to L(q) = l_1(p_1)+l_2(p_2)\, ,\qquad
W^\pm \to L(q) = l^\pm(p_1) + \nu(p_2) \, ,
\end{eqnarray}
and $p_{1,2}$ are the lepton momenta, $q=p_1+p_2$. 
In the presence of hadronic final states $X(k_i)$ 
from additional real emission radiation in eq.~(\ref{eq:V}), 
using $q = p_a+p_b-\sum_i k_i$, the gauge boson momentum can be expressed through 
$Q$, $Y$ and a non-vanishing transverse momentum $q_T$, such that in components 
\begin{eqnarray}
\label{eq:q-para}
q^\mu &=& \left(m_T\,\cosh(Y), q_T, 0, m_T\,\sinh(Y) \right)
\, ,
\nonumber \\
p_1^\mu &=& p_{T1} \left(\cosh (Y+\Delta y), \cos \phi, \sin \phi, \sinh (Y+\Delta y) \right)
\, ,
\nonumber \\
p_2^\mu &=& q^\mu - p_1^\mu 
\, ,
\end{eqnarray}
where $m_T = \sqrt{Q^2+q_T^2}$ and 
the azimuthal angle $\phi$ in the transverse plane is given by 
$\vec{p}_{T1} \cdot \vec{q}_T = p_{T1}\, q_T\, \cos \phi$.
Momentum conservation yields for the transverse momenta and rapidities of the leptons 
\begin{eqnarray}
\label{eq:pT-constraints}
p_{T1} &=& \frac{Q^2/2}{m_T\,\cosh(\Delta y) - q_T \cos\phi }
\, ,
\nonumber \\
p_{T2} &=& \sqrt{\left(p_{T1}\right)^2 - 2 p_{T1} q_T\cos\phi + q_T^2}
\, ,
\end{eqnarray}
and 
\begin{eqnarray}
\label{eq:eta-constraints}
\eta_{1} &=& Y+\Delta y
\, ,
\nonumber \\
\eta_{2} &=& Y+\frac{1}{2} \ln\left(\frac{m_T - p_{T1} \exp(+\Delta y)}{m_T - p_{T1} \exp(-\Delta y)}\right)
\, .
\end{eqnarray}

The leptonic final state phase space $\Phi_{L}$ (neglecting lepton masses) reads
in terms of the variables $\phi$ and $\Delta y$ in eq.~(\ref{eq:q-para}), 
\begin{eqnarray}
\label{eq:decay-phase-space}
\Phi_{L}(q_T) &=& 
\left( \int \prod_{i=1}^2\, \frac{d^4p_i}{(2\,\pi)^{3}}\, \delta^+(p_i^2)\,\right)\, (2\,\pi)^4\, \delta^{(4)}(q-p_1-p_2)
\,=\, 
\frac{1}{4\,\pi^2}\, 
\int_0^\pi d\phi\, \int_{-\infty}^{\infty} d\Delta y\, \frac{p_{T1}^2}{Q^2}\,  
 \, ,
\end{eqnarray}
and $p_{T1}$ implicitly depends on $\phi$ and $\Delta y$ through eq.~(\ref{eq:pT-constraints}).
The fiducial cuts on the decay leptons applied to the data discussed in Sec.~\ref{sec:benchmark} 
modify eq.~(\ref{eq:decay-phase-space}) by constraining the integration range. 
With the typical cuts on the transverse momenta and rapidities of the leptons, 
$p_{T1,2} \geq p_T^{\rm min}$ and $\eta_{1,2}^{\rm min} \leq \eta_{1,2} \leq \eta_{1,2}^{\rm max}$, 
the phase space $\Phi_{L}$ becomes
\begin{eqnarray}
\label{eq:phase-space-cuts}
\Phi_{L}(q_T) &=& 
\frac{1}{4\,\pi^2}\, 
\int_0^\pi d\phi\, \int_{-\infty}^{\infty} d\Delta y\, \frac{p_{T1}^2}{Q^2}\,  
\left( \prod_{i=1}^2\, \theta(p_{Ti}-p_T^{\rm min})\, 
\theta(\eta_i - \eta_i^{\rm min} ) \theta(\eta_i^{\rm max} - \eta_i) \right)
\, .
\end{eqnarray}

It has been pointed out in \cite{Ebert:2019zkb} that the presence 
of cuts on the leptons' transverse momenta 
breaks azimuthal symmetry and leads to linear power corrections in $q_T$. 
The expansion of eqs.~(\ref{eq:pT-constraints}) and (\ref{eq:eta-constraints}) for
small $q_T$ up to quadratic corrections in $q_T$ gives
\begin{eqnarray}
\label{eq:pT-constraints-exp}
p_{T1} &=& \frac{Q}{2\cosh(\Delta y)} + q_T \frac{\cos\phi}{2 \cosh^2(\Delta y)} + {\cal O}(q_T^2/Q) 
\, ,
\nonumber \\
p_{T2} &=& p_{T1} - q_T\cos\phi + {\cal O}(q_T^2/Q) 
\, ,
\end{eqnarray}
and 
\begin{eqnarray}
\label{eq:eta-constraints-exp}
\eta_{1} &=& Y+\Delta y
\, ,
\nonumber \\
\eta_{2} &=& Y - \Delta y - 2 \frac{q_T}{Q} \cos\phi \sinh(\Delta y) + {\cal O}(q_T^2/Q^2) 
\, .
\end{eqnarray}

The symmetric cut on the transverse momenta $p_{T1,2} \geq p_T^{\rm min}$ 
used for the ATLAS or D{\O} data considered above thus splits the $\phi$
integration range depending on the sign of $\cos \phi$ as 
\begin{eqnarray}
\label{eq:pT-min-cases}
{\rm min}\left\{p_{T1}, p_{T2}\right\} 
&=& \left\{
  \begin{tabular}{lcl}
    $p_{T1}\, ,$  & $\qquad$ & $\cos \phi < 0$ \\
    \\
    $p_{T1} - q_T \cos\phi\, ,$ & $\qquad$ & $\cos \phi \geq 0$     
  \end{tabular}
\right.
 \, ,
\end{eqnarray}
such that the $\theta$-functions in eq.~(\ref{eq:phase-space-cuts})
give rise to different integrands in the respective regions and 
the linear power corrections in $q_T$ do not vanish in the phase space integral.
Considering only the cut $p_{T1,2} \geq p_T^{\rm min}$ on the transverse momenta, the
phase space $\Phi_{L}$ becomes
\begin{eqnarray}
\label{eq:phase-space-pt-cuts}
\Phi_{L}(q_T) &=& 
\frac{1}{8\,\pi}\, \sqrt{1-(2 p_T^{\rm min})^2/Q^2}
- \frac{1}{2\,\pi^2}\,\frac{q_T}{Q}\, \frac{p_T^{\rm min} }{Q\,\sqrt{1-(2 p_T^{\rm min})^2/Q^2}}
+ {\cal O}(q_T^2/Q^2) 
\, .
\end{eqnarray}

In case of rapidity cuts, the constraints are slightly more involved, see \cite{Ebert:2020dfc}.
Considering the cuts $|\eta_{1,2}| \leq \eta^{\rm max}$ for the selection of both leptons at central pseudo-rapidity, 
as applied to the ATLAS data for $Z/\gamma^*$-boson production, the phase space $\Phi_{L}$ at Born level becomes
\begin{eqnarray}
\label{eq:phase-space-eta-cuts}
\Phi_{L}(q_T) &=& 
\frac{1}{8\,\pi}\, \tanh(\eta^{\rm max}-Y) + {\cal O}(q_T/Q) 
\, .
\end{eqnarray}
For small $Y$ this constraint is less tight compared to the $p_{T1,2}$ cuts in 
eq.~(\ref{eq:phase-space-pt-cuts}) for the typical values of $p_T^{\rm min}$ used.
To assess where the $\theta$-functions in eq.~(\ref{eq:phase-space-cuts}) involving the 
pseudo-rapidities affect the phase space $\Phi_{L}$ and, in particular, 
break azimuthal symmetry, one has to examine the regions where they overlap.
The boundary of this region is given by $|\eta_1| = |\eta_2|$, 
which determines a value $\phi^\ast$ through the condition
\begin{eqnarray}
\label{eq:cosphi-limit}
\cos \phi^\ast &=& \frac{Q}{2 q_T}\, \frac{\sinh(2Y)} {\sinh(2Y+\Delta y)} + {\cal O}(q_T/Q) 
\, .
\end{eqnarray}
The constraint $|\cos \phi^\ast| \leq 1$ restricts either the region of $|\eta_1|$ or $|\eta_2|$ in the phase space (\ref{eq:phase-space-cuts}), but not both. 
Thus, azimuthal symmetry is not broken by the pseudo-rapidity cuts, 
but may still be affected by $p_T$ cuts through eq.~(\ref{eq:pT-min-cases}). 
Solutions in the physical range for $|\cos \phi^\ast| < 1$ imply the following 
scaling for the values of $q_T/Q$
\begin{eqnarray}
\label{eq:qT-limit}
\frac{q_T}{Q} &>& \frac{1}{2}\, \left| \frac{\sinh(2Y)} {\sinh(2Y+\Delta y)} \right| 
\, ,
\end{eqnarray}
which can be approximated for small $Y$ (and using $|\eta_{1}| = |Y+\Delta y| \leq \eta^{\rm max}$)

\begin{eqnarray}
\label{eq:qT-limit-exp}
\frac{q_T}{Q} &\gtrsim & \frac{q^\ast_T}{Q} \,=\, \frac{|Y|} {\sinh(\eta^{\rm max})} +  {\cal O}(Y) 
\, .
\end{eqnarray}
This defines a lower bound on $q_T$ for linear power corrections to appear as a result
of broken azimuthal symmetry due to the pseudo-rapidity cuts. 
For values $q_T < q^\ast_T$ azimuthal symmetry is restored and only quadratic power corrections arise. 
As eqs.~(\ref{eq:cosphi-limit})-(\ref{eq:qT-limit-exp}) indicate, the
transition between these two regions of $q_T$ is sharp up to corrections.

\begin{figure}[t!]
\begin{center}
\includegraphics[width=8.1cm]{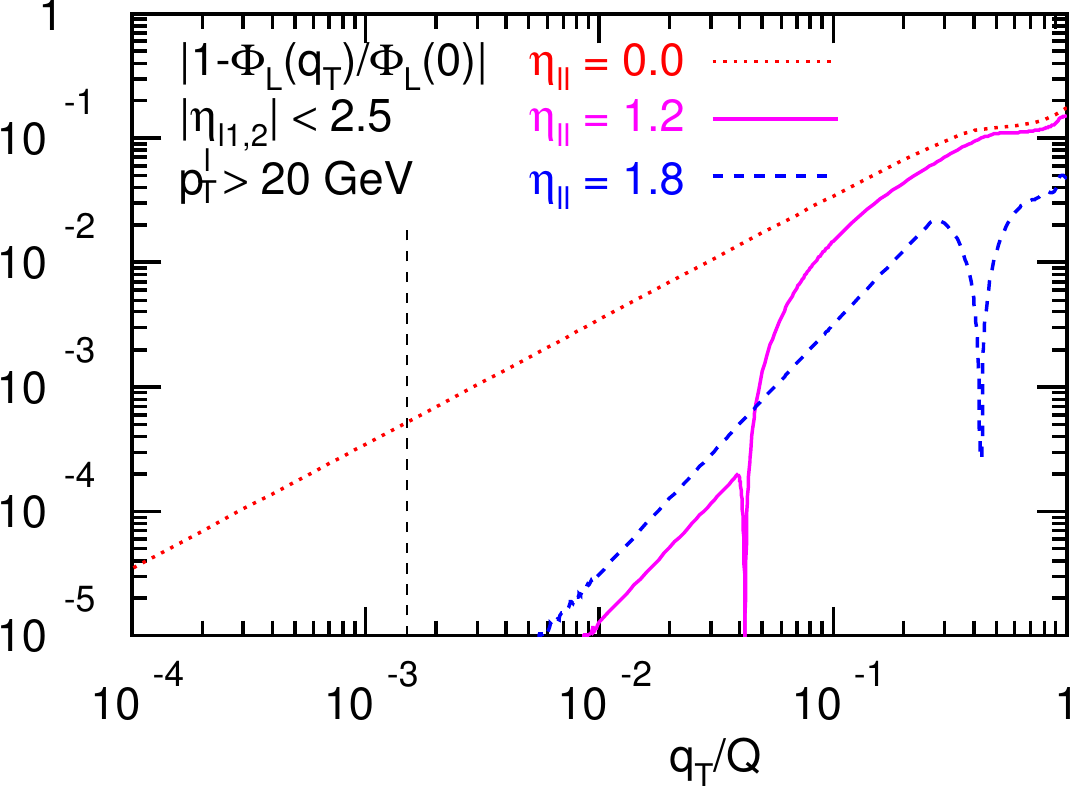}
\includegraphics[width=8.1cm]{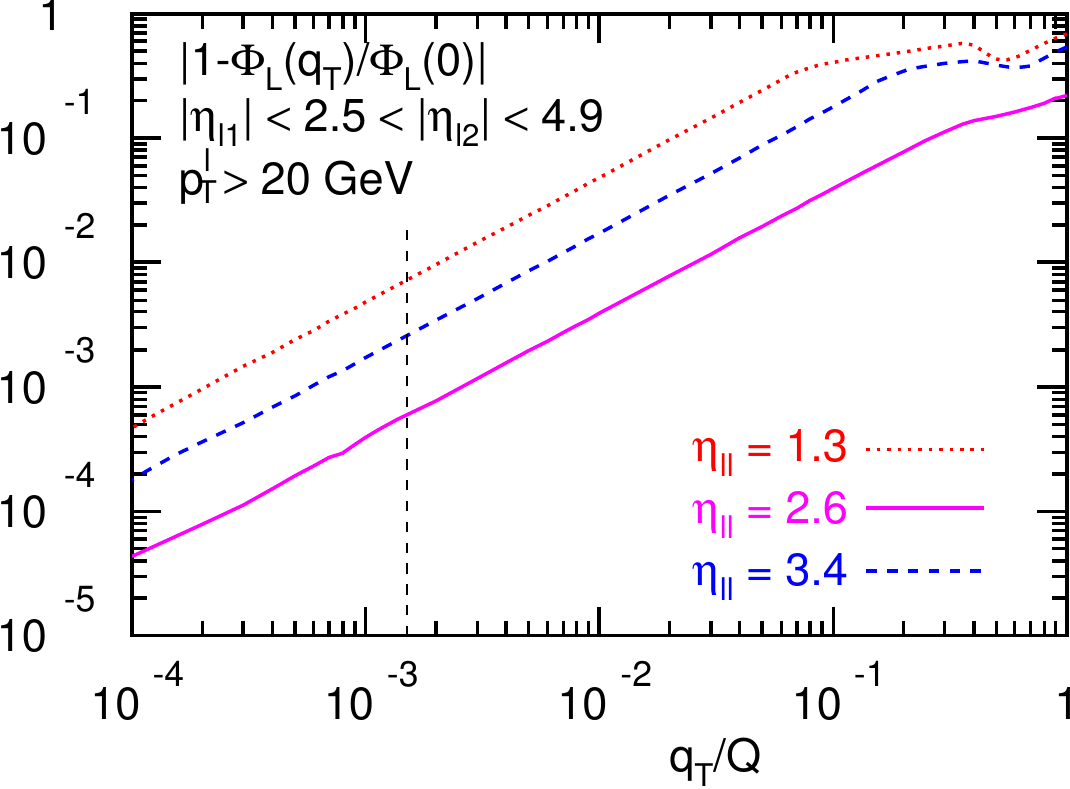}
\caption{\small
  \label{fig:dPhi-atlas-Z-boson}
The difference between the Born and real emission phase spaces $\Phi_L(0)-\Phi_L(q_T)$ of the decay leptons relative to the Born one
at fiducial cuts applied to 
ATLAS data set~\cite{Aaboud:2016btc} for $Z/\gamma^*$-boson production
($Q=M_Z$) for different values of the gauge boson pseudo-rapidity $\eta_{ll}$.
For the lepton momenta $p_T^l \geq 20$~GeV are required.
Left: Cuts selecting central pseudo-rapidities $|\eta_{l_i}| \leq 2.5$ for $i=1,2$.
Right: Cuts selecting one lepton at central
pseudo-rapidity $|\eta_{l_1}| \leq 2.5$ and the other 
at forward pseudo-rapidity, $2.5 \leq |\eta_{l_2}| \leq 4.9$. 
The vertical dashed line indicates the minimum value $r^{\rm min}_{\rm cut}=0.15\%$ 
used in {\tt MATRIX} as a slicing cut.
}
\end{center}
\end{figure}
\begin{figure}[t!]
\begin{center}
\includegraphics[width=8.1cm]{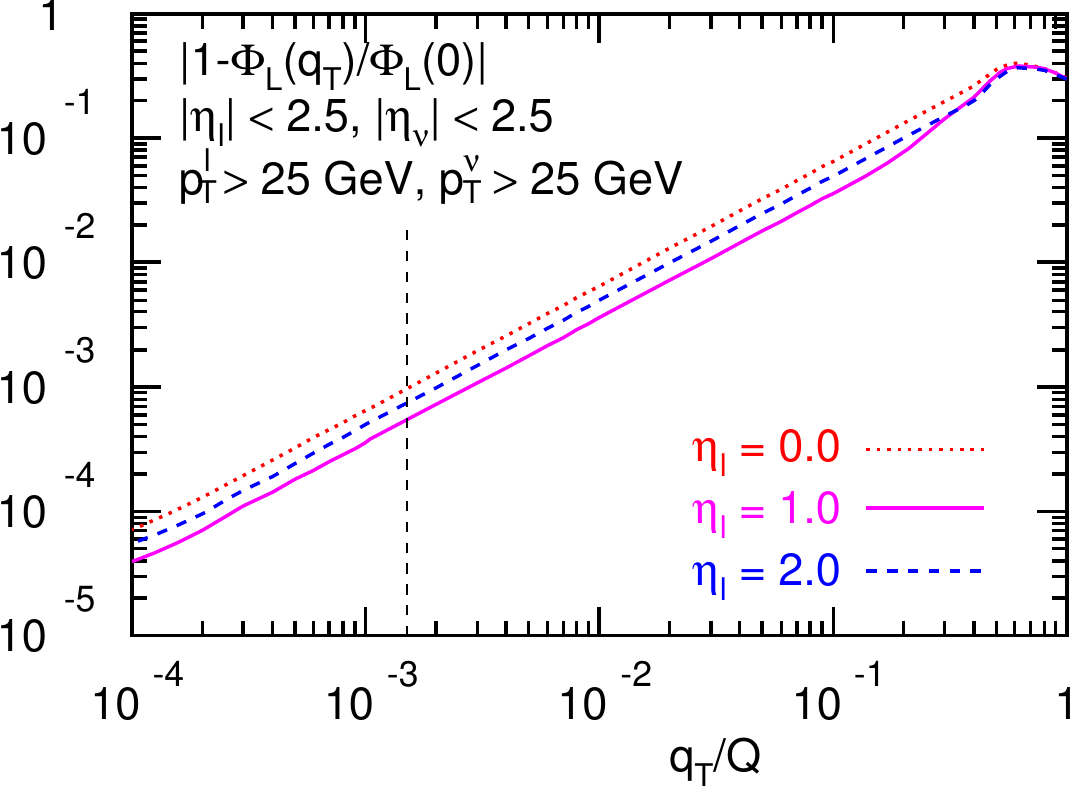}
\caption{\small
  \label{fig:dPhi-W-boson}
Same as Fig.~\ref{fig:dPhi-atlas-Z-boson} for the fiducial cuts applied to 
ATLAS data set~\cite{Aaboud:2016btc} for $W^\pm$-boson production ($Q=M_W$) and different
values of the lepton pseudo-rapidity $\eta_l$. 
For the lepton momenta $p_T^{l,\nu} \geq 25$~GeV are required.
}
\end{center}
\end{figure}

The appearance of linear power corrections in the phase space $\Phi_{L}$ 
can be illustrated by considering the deviations 
$\left| 1 - \Phi_{L}(q_T)/\Phi_{L}(0)\right|$ from the Born level leading power results for $q_T=0$.
In Fig.~\ref{fig:dPhi-atlas-Z-boson} we show them for the fiducial cuts
applied to the ATLAS data in case of $Z/\gamma^*$-boson production.
On the left in Fig.~\ref{fig:dPhi-atlas-Z-boson}, the leptons are selected 
at central pseudo-rapidities $|\eta_{l_i}| \leq 2.5$ for $i=1,2$ and we
observe the presence of linear power corrections in $q_T$ 
for central gauge boson pseudo-rapidities $\eta_{ll} \lesssim 1$ 
due to the $p_T$ constraint in eq.~(\ref{eq:pT-min-cases}).
For $r_{\rm cut}=q_T/Q=0.15\%$, which is the default value for $r_{\rm cut}$ 
used in {\tt MATRIX} as a slicing cut and indicated by the vertical dashed line in Fig.~\ref{fig:dPhi-atlas-Z-boson}, 
their size amounts to ${\cal  O}(0.5\permille)$.
In contrast, for larger $\eta_{ll}$ the pseudo-rapidity constraints dominate the 
phase space $\Phi_{L}$ and azimuthal symmetry is restored, resulting in 
quadratic power corrections in $q_T$ for small enough $q_T$, 
see eqs.~(\ref{eq:cosphi-limit})-(\ref{eq:qT-limit-exp}).
In Fig.~\ref{fig:dPhi-atlas-Z-boson} on the left this feature is illustrated
for $\eta_{ll} = 1.2$ and $1.8$, and the corrections to $\Phi_{L}$ for $r_{\rm cut}=0.15\%$ 
are smaller by more than two orders of magnitude.
It is interesting to compare these findings with the $\eta_{ll}$ dependence of
the differences at NLO of {\tt DYNNLO} from codes using local subtraction 
in Fig.~\ref{fig:atlas-Z-central-nlo} and with the deviations at NNLO of  
{\tt DYNNLO}, {\tt MATRIX} (with $r_{\rm cut}=0.15\%$) and {\tt MCFM} from {\tt FEWZ} in Figs.~\ref{fig:dynnlo-atlas7data}, \ref{fig:matrix-atlas7data} and \ref{fig:mcfm-atlas7data} for central $Z/\gamma^*$-boson production.
For $\eta_{ll} \lesssim 1$ all slicing codes undershoot {\tt FEWZ}, while they 
tend to agree well for $\eta_{ll} \gtrsim 1.5$. 
The transition around  $\eta_{ll} \gtrsim 1.2$ when the linear power corrections in $q_T$ in $\Phi_L$ vanish, 
is most pronounced in the case of {\tt MCFM} in Fig.~\ref{fig:mcfm-atlas7data}.

In Fig.~\ref{fig:dPhi-atlas-Z-boson} on the right we plot the same study for 
the ATLAS cuts with one lepton at central and the other at forward pseudo-rapidity. 
In this case, due to the non-overlapping regions 
$|\eta_{l_1}| \leq 2.5$ and $2.5 \leq |\eta_{l_2}| \leq 4.9$ 
azimuthal symmetry is always broken by the $p_T$ constraint in eq.~(\ref{eq:pT-min-cases}) 
and we observe sizable linear power corrections. 
For the chosen values of $\eta_{ll}=1.3$, 2.6 and $3.4$ they amount to 
${\cal  O}(7\permille)$, ${\cal O}(0.6\permille)$ and ${\cal  O}(2\permille)$ at the value of $r_{\rm cut}=0.15\%$ in {\tt MATRIX},  
and the relatively large size of these corrections is remarkable.
Moreover, their $\eta_{ll}$ dependence matches well with the pattern of 
the observed deviations  
of {\tt DYNNLO} from codes with dipole subtraction at NLO in Fig.~\ref{fig:atlas-Z-forward-nlo} on the left 
and of {\tt DYNNLO}, {\tt MATRIX} and {\tt MCFM} from {\tt FEWZ} at NNLO
in Figs.~\ref{fig:dynnlo-atlas7data}, \ref{fig:matrix-atlas7data} and \ref{fig:mcfm-atlas7data}, 
where the {\tt DYNNLO}, {\tt MATRIX} and {\tt MCFM} deliver significantly
smaller results in the first and the last $\eta_{ll}$ bins.

In Fig.~\ref{fig:dPhi-W-boson} we plot the phase space $\Phi_{L}$ for 
the fiducial cuts applied to the ATLAS data set~\cite{Aaboud:2016btc} for $W^\pm$-boson production.
In this case, the binning in the lepton pseudo-rapidity fixes $\eta_l$ and
trivially fulfills one of the $\theta$-functions in the integral for $\Phi_{L}$ in eq.~(\ref{eq:phase-space-cuts}).
The other $\theta$-function constrains the neutrino's pseudo-rapidity in the whole integration range, 
so that the pseudo-rapidity cuts do not restore the azimuthal symmetry for small $q_T$.
The linear power corrections in $q_T$, which we observe in Fig.~\ref{fig:dPhi-W-boson} 
originate from the constraints $p_T^{l,\nu} \geq 25$~GeV for the lepton momenta, 
which do break azimuthal symmetry as shown in eq.~(\ref{eq:pT-min-cases}).
They amount to corrections ${\cal  O}(0.4-0.8\permille)$ for the value $r_{\rm cut}=0.15\%$, 
depending on the value of the lepton pseudo-rapidity $\eta_l$, 
with larger corrections observed for central $\eta_l$. 
This pattern is in line with the observed deviations 
in Fig.~\ref{fig:atlas-Wpm-nlo} at NLO between {\tt DYNNLO} and local subtraction results, and 
in Figs.~\ref{fig:dynnlo-atlas7data}, \ref{fig:matrix-atlas7data} and \ref{fig:mcfm-atlas7data} 
at NNLO between {\tt DYNNLO}, {\tt MATRIX} and {\tt MCFM} on the one and {\tt FEWZ} on the
other side, where all slicing codes give consistently lower results than {\tt FEWZ} 
and the deviations display little dependence on the lepton pseudo-rapidity $\eta_l$.

\begin{figure}[t!]
\begin{center}
\includegraphics[width=8.1cm]{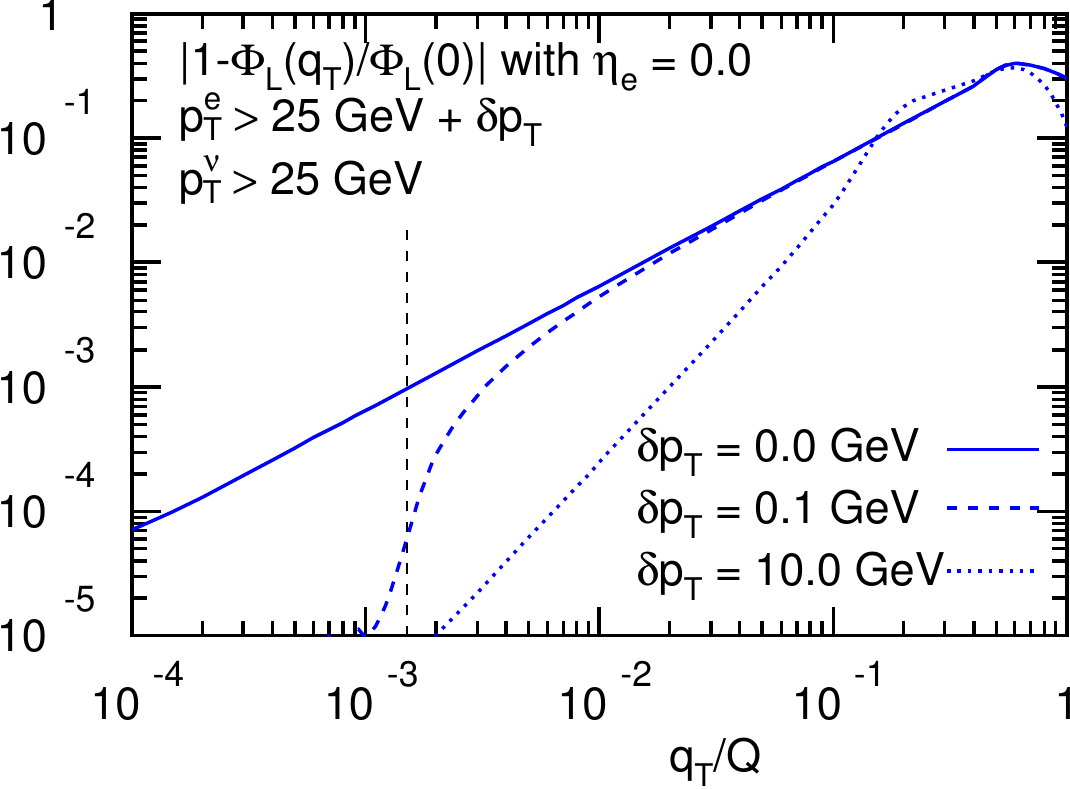}
\caption{\small
  \label{fig:dPhi-W-boson-D0-with-Delta}
Same as Fig.~\ref{fig:dPhi-atlas-Z-boson} for the fiducial cuts applied to the D{\O} data~\cite{D0:2014kma}
for the electron charge asymmetry distribution $A_e$, using staggered cuts 
$p_{T1} \geq p_T^{\rm min}+\delta p_{T}$, $p_{T2} \geq p_T^{\rm min}$ and $Q=M_W$ 
for the electron pseudo-rapidity fixed at $\eta_{e}=0$.
}
\end{center}
\end{figure}
Finally, we briefly discuss staggered cuts on the transverse momenta, when 
$p_{T1} \geq p_T^{\rm min}+\delta p_{T}$ and $p_{T2} \geq p_T^{\rm min}$ for some $\delta p_{T} > 0$, 
as realized for instance in the D{\O} measurement of the electron charge
asymmetry distribution $A_e$ discussed above.
For staggered cuts with $\delta p_{T}$ the phase space $\Phi_{L}$ evaluates at
Born level then as 
\begin{eqnarray}
\label{eq:phase-space-pt-cuts-with-Delta}
\Phi_{L}(q_T) &=& 
\frac{1}{8\,\pi}\, \sqrt{1- (2 (p_T^{\rm min}+\delta p_T))^2/Q^2} + {\cal O}(q_T/Q) 
\, .
\end{eqnarray}
In addition, with staggered cuts the $\theta$-functions in eq.~(\ref{eq:phase-space-cuts}) 
which constrain the fiducial phase space give rise to the following 
condition for the azimuthal integration
\begin{eqnarray}
\label{eq:pT-min-cases-with-Delta}
{\rm min}\left\{p_{T1} - \delta p_{T}, p_{T2}\right\} 
&=& \left\{
  \begin{tabular}{lcl}
    $p_{T1}\, - \delta p_{T},$  & $\qquad$ & $\cos \phi < \delta p_{T}/q_T$ \\
    \\
    $p_{T1} - q_T \cos\phi\, ,$ & $\qquad$ & $\cos \phi \geq \delta p_{T}/q_T$     
  \end{tabular}
\right.
 \, ,
\end{eqnarray}
which reproduces the case of symmetric cuts in eq.~(\ref{eq:pT-min-cases}) for $\delta p_{T} \to 0$. 
On the other hand, it is obvious that for $\delta p_{T} > q_T$ 
the $\theta$-functions for the cuts on the transverse momenta 
in eq.~(\ref{eq:phase-space-cuts}) have no regions of common overlap, 
and therefore, do not affect the azimuthal symmetry, which is unbroken in this case.
The typical scales for the slicing cuts $r_{\rm cut}$ or $\tau_{\rm cut}$ 
in the slicing codes imply that $q_T^{\rm min} \ll 1$\,GeV, 
while measurements with staggered cuts in the experiments use 
values of $\delta p_{T}$ of the order of a few GeV.
Therefore, due to unbroken azimuthal symmetry, linear power corrections are largely absent.

We illustrate the effect of the staggered cuts in Fig.~\ref{fig:dPhi-W-boson-D0-with-Delta}
for the fiducial cuts applied to the D{\O} data~\cite{D0:2014kma} 
on the electron charge asymmetry distribution $A_e$.
We apply a series of cuts $\delta p_{T} = 0$, $0.1$ and $10$~GeV 
for an electron pseudo-rapidity of $\eta_{e}=0$. 
Other choices of $\eta_{e}$ give qualitatively the same results, see Fig.~\ref{fig:dPhi-W-boson}.
As the value of $\delta p_{T}$ increases, the deviations from the linear power
corrections for the case of $\delta p_{T} = 0$~GeV become apparent.
For $\delta p_{T}=10$~GeV, which corresponds to the choice in the D{\O}
data selection of $p_T^{e} \geq 35$~GeV and $p_T^{\nu} \geq 25$~GeV, we
observe in Fig.~\ref{fig:dPhi-W-boson} only quadratic power corrections 
for small $q_T$. This is in line with the general good agreement between the
results of {\tt DYNNLO}, {\tt MATRIX}, {\tt MCFM} and {\tt FEWZ} observed in 
Figs.~\ref{fig:dynnlo-d0-staggered} and \ref{fig:mcfm-d0-staggered}.
In addition, the observed pattern for the convergence of slicing codes in the limit of
vanishing slicing cuts $r_{\rm cut}$ or $\tau_{\rm cut}$ 
also agrees with studies performed in the presence of staggered cuts with 
{\tt MATRIX} in~\cite{Grazzini:2017mhc} and with {\tt MCFM} in~\cite{Campbell:2019dru}.

In summary, the changes in the relative size of the power corrections in $\Phi_{L}$ 
are correlated with better or worse agreement between the cross
sections generated with phase space slicing codes ({\tt DYNNLO}, {\tt MATRIX} and {\tt MCFM})
and the one with a local subtraction ({\tt FEWZ}).
This holds in particular for the dependence on the gauge boson pseudo-rapidity $\eta_{ll}$
in the case of $Z/\gamma^*$-boson production and the effect of staggered cuts.
This observation does not prove that the neglected power corrections are the pure source of the differences among the predictions of the various codes. Nevertheless, it is at least a warning sign that the computation of the NNLO corrections can only be considered a solved problem if the numerical precision of the Monte Carlo integration is under better control than the size of the correction itself, which calls for an improvement of the presently available codes and/or subtraction methods at NNLO accuracy.
The linear power corrections can be uniquely predicted by factorization.
This fact has been used in~\cite{Ebert:2020dfc} to propose a method for the inclusion of all fiducial-cut induced power corrections in
the framework of $q_T$-subtraction schemes 
and has been applied to provide cross section predictions for Higgs-boson production with fiducial cuts~\cite{Billis:2021ecs}.

\section{Conclusions}
\label{sec:conclusions}

We have investigated the accuracy of available NNLO QCD predictions 
for the hadroproduction of $W^{\pm}$- and $Z$-bosons, including their leptonic decays 
and keeping fully exclusive kinematics. 
Such predictions are available from several publicly available codes and we
have chosen {\tt DYNNLO}, {\tt  FEWZ}, {\tt MATRIX} and {\tt MCFM} to compute
benchmark values for kinematics which are representative for 
measurements of differential distributions 
in the pseudo-rapidities of the decay leptons from the LHC and Tevatron. 
The uncertainties in the cross sections from the numerical Monte Carlo
integration have been limited to few units in $10^{-4}$ and are negligible in
all cases.
At NLO there is perfect agreement among the results from {\tt  FEWZ}, {\tt MATRIX} and 
{\tt MCFM} and, partially with those of {\tt DYNNLO} as well. 
However, at NNLO accuracy we found differences among the predictions by the same codes that are comparable to the NNLO correction itself.
We demonstrated that the observed systematic differences can be understood in terms of 
the subtraction schemes employed, {\tt  FEWZ} using a fully local subtraction
scheme and {\tt DYNNLO}, {\tt MATRIX} and {\tt MCFM} applying phase space slicing schemes.
We have illustrated, how the fiducial cuts on the transverse momenta and 
pseudo-rapidities of the decay leptons lead to linear power corrections in the slicing parameter, 
i.e.~$q_T/Q$ and $\sqrt{{\cal T}_0/Q}$ for the $q_T$ and $N$-jettiness subtractions. 
Also the deviations share certain patterns across the range of
pseudo-rapidities in the considered distributions, which have been 
correlated with the appearance of linear power corrections 
in the lepton decay phase space $\Phi_{L}$ as a function of $q_T$.
The latter serves as a simple and efficient model to study 
power corrections in cross sections for the gauge boson production
with their subsequent leptonic decay.
For most of the distributions considered the pure NNLO QCD corrections on top
of the NLO ones are rather small, often in the range of ${\cal O}(1\%)$, 
while the deviations among the codes investigated in this study are not
substantially smaller, often even of the same size or larger, hinting towards 
a significant intrinsic uncertainty in the computation of the NNLO QCD
corrections for those observables.

In summary, with the continuous increase in the precision of the experimental measurements,
the theory predictions are pressed to provide cross sections at NNLO (or 
beyond) where the systematic uncertainties due to choices of particular
schemes or algorithms for the computation can be safely neglected in
comparison to the experimental uncertainties. 
The results of our study call for further improvements in this direction.

\subsection*{Acknowledgments}
We are grateful to Tobias Neumann for useful discussions on {\tt MCFM}.

This article is based upon work from COST Action CA16201
PARTICLEFACE supported by COST (European Cooperation in Science
and Technology).
The work of S.A. is supported by the DFG grants MO 1801/5-1, KN 365/14-1. 
S.M. acknowledges support by the {\it MTA Distinguished Guest Scientists Fellowship Programme in Hungary}. This research was supported in part by the National Science Foundation under Grant No. NSF PHY-1748958
and by grant K 125105 of the National Research, Development and Innovation Fund in Hungary.


\begin{thebibliography}{10}

\bibitem{Heinrich:2020ybq}
G.~Heinrich,
\newblock Phys. Rept. \textbf{922}, 1 (2021), arXiv:2009.00516.

\bibitem{Catani:2007vq}
S.~Catani and M.~Grazzini,
\newblock Phys. Rev. Lett. {\bf 98}, 222002 (2007), arXiv:hep-ph/0703012.

\bibitem{Catani:2009sm}
S.~Catani {\em et~al.},
\newblock Phys. Rev. Lett. {\bf 103}, 082001 (2009), arXiv:0903.2120.

\bibitem{Gavin:2010az}
R.~Gavin, Y.~Li, F.~Petriello, and S.~Quackenbush,
\newblock Comput. Phys. Commun. {\bf 182}, 2388 (2011), arXiv:1011.3540.

\bibitem{Boughezal:2016wmq}
R.~Boughezal {\em et~al.},
\newblock Eur. Phys. J. {\bf C77}, 7 (2017), arXiv:1605.08011.

\bibitem{Aaboud:2016btc}
ATLAS, M.~Aaboud {\em et~al.},
\newblock Eur. Phys. J. C {\bf 77}, 367 (2017), arXiv:1612.03016.

\bibitem{Alioli:2016fum}
S.~Alioli {\em et~al.},
\newblock Eur. Phys. J. {\bf C77}, 280 (2017), arXiv:1606.02330.

\bibitem{D0:2014kma}
D0, V.~M. Abazov {\em et~al.},
\newblock Phys. Rev. {\bf D91}, 032007 (2015), arXiv:1412.2862,
\newblock [Erratum: Phys. Rev.D91,no.7,079901(2015)].

\bibitem{Aad:2019rou}
ATLAS, G.~Aad {\em et~al.},
\newblock Eur. Phys. J. {\bf C79}, 760 (2019), arXiv:1904.05631.

\bibitem{Khachatryan:2016pev}
CMS, V.~Khachatryan {\em et~al.},
\newblock Eur. Phys. J. {\bf C76} 469 (2016), arXiv:1603.01803.

\bibitem{Aaij:2015zlq}
LHCb, R.~Aaij {\em et~al.}, 
\newblock JHEP {\bf 01}, 155 (2016), arXiv:1511.08039.

\bibitem{Dittmaier:2009cr}
S.~Dittmaier and M.~Huber,
\newblock JHEP {\bf 01}, 060 (2010), arXiv:0911.2329.

\bibitem{pdg2020}
Particle Data Group, P.~A. Zyla {\em et~al.},
\newblock Progress of Theoretical and Experimental Physics {\bf 2020} (2020),
\newblock 083C01.

\bibitem{Alekhin:2017kpj}
S.~Alekhin, J.~Bl{\"u}mlein, S.~Moch, and R.~Placakyte,
\newblock Phys. Rev. {\bf D96}, 014011 (2017), arXiv:1701.05838.

\bibitem{Alekhin:2018pai}
S.~Alekhin, J.~Bl{\"u}mlein, and S.~Moch,
\newblock Eur. Phys. J. {\bf C78}, 477 (2018), arXiv:1803.07537.

\bibitem{Li:2012wna}
Y.~Li and F.~Petriello,
\newblock Phys. Rev. {\bf D86}, 094034 (2012), arXiv:1208.5967.

\bibitem{Gavin:2012sy}
R.~Gavin, Y.~Li, F.~Petriello, and S.~Quackenbush,
\newblock Comput. Phys. Commun. {\bf 184}, 208 (2013), arXiv:1201.5896.

\bibitem{Grazzini:2017mhc}
M.~Grazzini, S.~Kallweit, and M.~Wiesemann,
\newblock Eur. Phys. J. {\bf C78}, 537 (2018), arXiv:1711.06631.

\bibitem{Cascioli:2011va}
F.~Cascioli, P.~Maierhofer, and S.~Pozzorini,
\newblock Phys. Rev. Lett. {\bf 108}, 111601 (2012), arXiv:1111.5206.

\bibitem{Campbell:2019dru}
J.~Campbell and T.~Neumann,
\newblock JHEP {\bf 12}, 034 (2019), arXiv:1909.09117.

\bibitem{Boughezal:2015dva}
R.~Boughezal, C.~Focke, X.~Liu, and F.~Petriello,
\newblock Phys. Rev. Lett. {\bf 115}, 062002 (2015), arXiv:1504.02131.

\bibitem{Gaunt:2015pea}
J.~Gaunt, M.~Stahlhofen, F.~J. Tackmann, and J.~R. Walsh,
\newblock JHEP {\bf 09}, 058 (2015), arXiv:1505.04794.

\bibitem{TorresBobadilla:2020ekr}
W.~J. Torres~Bobadilla {\em et~al.},
\newblock Eur. Phys. J. C {\bf 81}, 250 (2021), arXiv:2012.02567.

\bibitem{Camarda:2019zyx}
S.~Camarda {\em et~al.},
\newblock Eur. Phys. J. C {\bf 80}, 251 (2020), arXiv:1910.07049,
\newblock [Erratum: Eur.Phys.J.C 80, 440 (2020)].

\bibitem{Hoeche:2014aia}
S.~H{\"o}che, Y.~Li, and S.~Prestel,
\newblock Phys. Rev. {\bf D91}, 074015 (2015), arXiv:1405.3607.

\bibitem{Catani:1996vz}
S.~Catani and M.~Seymour,
\newblock Nucl. Phys. B {\bf 485}, 291 (1997), arXiv:hep-ph/9605323,
\newblock [Erratum: Nucl.Phys.B 510, 503--504 (1998)].

\bibitem{Moult:2017jsg}
I.~Moult {\em et~al.},
\newblock Phys. Rev. D {\bf 97}, 014013 (2018), arXiv:1710.03227.

\bibitem{Boughezal:2018mvf}
R.~Boughezal, A.~Isgr{\`o}, and F.~Petriello,
\newblock Phys. Rev. D {\bf 97}, 076006 (2018), arXiv:1802.00456.

\bibitem{Ebert:2018gsn}
M.~A.~Ebert  {\em et~al.},
\newblock JHEP {\bf 04} 123 (2019), arXiv:1812.08189.

\bibitem{Cieri:2019tfv}
L.~Cieri, C.~Oleari and M.~Rocco,
\newblock Eur. Phys. J. C {\bf 79}, 852 (2019), arXiv:1906.09044.

\bibitem{Ebert:2019zkb}
M.~A. Ebert and F.~J. Tackmann,
\newblock JHEP {\bf 03}, 158 (2020), arXiv:1911.08486.

\bibitem{Moult:2016fqy}
I.~Moult {\em et~al.},
\newblock Phys. Rev. D {\bf 95}, 074023 (2017), arXiv:1612.00450.

\bibitem{Ebert:2018lzn}
M.~A. Ebert {\em et~al.},
\newblock JHEP {\bf 12}, 084 (2018), arXiv:1807.10764.

\bibitem{Ebert:2020dfc}
M.~A. Ebert, J.~K. Michel, I.~W. Stewart, and F.~J. Tackmann,
\newblock JHEP {\bf 04} 102 (2021), arXiv:2006.11382.

\bibitem{Billis:2021ecs}
G. Billis {\em et~al.},
\newblock (2021) arXiv:2102.08039.

\end{thebibliography}
\end{document}